\documentclass[a4paper,10pt]{article}
\usepackage[utf8]{inputenc}

\usepackage{authblk}
\usepackage{graphicx} 
\usepackage{balance} 
\usepackage{amssymb}

\usepackage{epstopdf}

\usepackage{multirow} 
\usepackage{array}

\usepackage{amsmath} 
\usepackage{wrapfig} 
\usepackage[utf8]{inputenc} 
\usepackage{clrscode} 
\usepackage{listings} 
\usepackage{subfigure} 
\usepackage{hyperref}
\usepackage{gensymb}
\usepackage{tikz}
\usepackage{verbatim}
\usepackage{scalefnt}
\usetikzlibrary{arrows}
\usetikzlibrary{positioning}
\usetikzlibrary{decorations.markings}
\usetikzlibrary{decorations.pathreplacing}

\title{On Quantifying Qualitative Geospatial Data: A Probabilistic Approach}

\author[1]{Georgios Skoumas\thanks{gskoumas@dblab.ece.ntua.gr}}
\author[2]{Dieter Pfoser\thanks{dpfoser@gmu.edu}}
\author[3]{Anastasios Kyrillidis \thanks{anastasios.kyrillidis@epfl.ch}}
\affil[1]{Knowledge and Database Systems Laboratory, National Technical University of Athens, Greece}
\affil[2]{Dept. of Geography and Geoinformation Science, George Mason University, USA}
\affil[3]{Laboratory for Information and Inference Systems, Ecole Polytechnique Federale de Lausanne, Switzerland}

\begin{document}

\maketitle

\begin{abstract}
Living in the era of data deluge, we have witnessed a web content explosion, largely due to the massive availability of User-Generated Content (UGC). 
In this work, we specifically consider the problem of geospatial information extraction and representation, 
where one can exploit diverse sources of information (such as image and audio data, text data, etc), 
going beyond traditional volunteered geographic information. 
Our ambition is to include available narrative information in an effort to better explain 
geospatial relationships: with spatial reasoning being a basic form of human cognition, 
narratives expressing such experiences typically contain qualitative spatial data, i.e., 
spatial objects and spatial relationships. 

To this end, we formulate a quantitative approach for the representation of qualitative spatial relations extracted from UGC in the form of texts. The proposed 
method quantifies such relations based on multiple text observations. Such observations provide distance and orientation features which 
are utilized by a greedy Expectation Maximization-based (EM) algorithm to infer a probability distribution over predefined spatial relationships; 
the latter represent the quantified relationships under user-defined probabilistic assumptions. We evaluate the applicability and quality of the proposed 
approach using real UGC data originating from an actual travel blog text corpus. To verify the result quality, we generate grid-based “maps” visualizing the spatial 
extent of the various relations.
\end{abstract}
%!TEX root = paper.tex
\section{Introduction} 
\label{sec:introduction}

During the last decade, we have witnessed an explosion in the amount and variety of content available on the Web. Sophisticated analysis of such UGC 
has become an important issue in many cutting edge research fields such as Geographical Information Science. In this work, our goal is to take advantage of such volunteered geographic information
in geospatial data analysis. In particular, when applied to the geospatial domain, this translates to massively collecting and sharing knowledge in order to ultimately model and chart the world.

Traditionally, quantitative information in the form of spatial coordinates is the data used in virtually 
all geospatial applications. With spatial reasoning being a basic form of human cognition, 
qualitative spatial data in the form of spatial relationships (North, South, In, Close, Next, Far, etc.)
%expressing, e.g., topology (inside, overlap), direction relations (e.g., North, South), ordinal relations (e.g., inside, contain), and distance relations (e.g., far, near), 
is what people typically use in order to describe spatial scenarios. 
Such data makes a prime source for user-contributed geospatial content. 
Especially narratives expressing such experiences typically contain spatial knowledge. 
However, one of the drawbacks of this data is its lack of precision as qualitative relations are interpreted differently by the users (in contrast to coordinates).

\begin{figure}[htbp]
\begin{center}
\includegraphics[width=4.3in,height=2.4in]{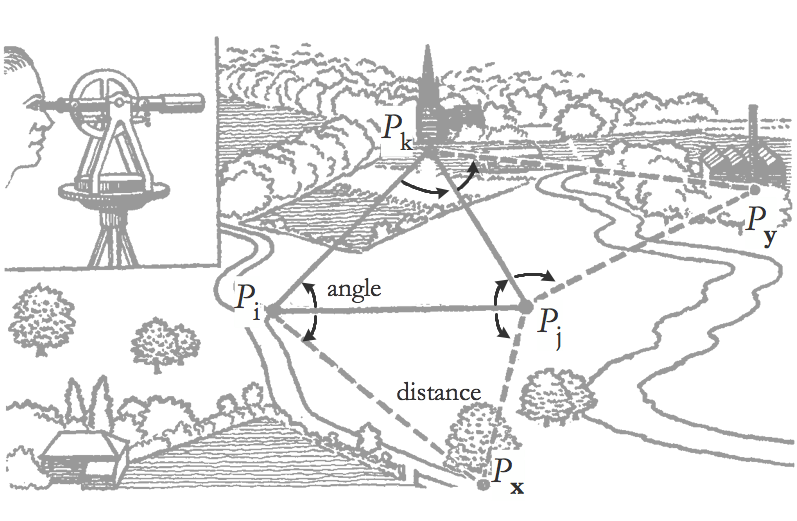} \caption{Intuitive problem formalization. $P_i$, $P_j$ and $P_k$ represent known locations that are used to compute several unknown locations ($P_x$ and $P_y$) based on distance and angle observations.} 
\end{center}
\label{fig:intuition}  
\end{figure}

As a motivational example we could consider the sentence; \textit{``Big Ben is the nickname for the great bell of the clock at the north end of the Palace of Westminster''}. 
In this case, we want to quantify what people imply when they say \textit{``North''} in terms of distance and direction. 
Having quantified \textit{``North''} in this context and knowing the location of either \textit{``Big Ben''} or \textit{``Palace of Westminster''} will allow us to infer 
possible locations for the other. Eventually, by collecting more observations of this form, we will be able to refine the location and, thus, locate spatial objects that otherwise could not be geocoded. 
Figure~\ref{fig:intuition} illustrates the underlying idea by relating our observation-based approach to the triangulation problem from surveying engineering, where an unknown 
location is determined by ``observing'' known locations. 

% \begin{figure}[htbp]
% \includegraphics[width=\columnwidth]{drawings/intuition.png} \caption{Intuitive Problem Formalization} 
% \label{fig:intuition} 
% \end{figure}

To this end, we consider the following problem:
\vskip.05in
\noindent \textsc{Problem:}
%\textit{Considering a set of spatial objects (Point of interests) $P = \{ P_1, P_2, \ldots, P_m \}$ whose positions in space is known, a set of spatial objects $U = \{U_1, U_2, \ldots, U_n \}$ whose positions in space is unknown, and a set of spatial relationships \\$R = \{R_1, R_2, \ldots, R_o\}$, find a probabilistic estimate of the positions of $U$ based on their spatial relationship $R$ with $O$. }
\textit{Given a set of spatial objects $\mathcal{K}$ whose positions in space are known, a set of spatial objects $\mathcal{U}$ whose 
positions in space are unknown, and a set of spatial relationships $\mathcal{R}$, find probabilistic estimates of the positions 
of objects of set $\mathcal{U}$ based on their spatial relationships $\mathcal{R}$ with objects of set $\mathcal{K}$.} 
%A relationship instance is of the form $u \bigodot p$, with $u \in \mathcal{U}$, $p \in \mathcal{P}$ and $\bigodot \in \mathcal{R}$.}
\vskip.05in

To achieve this, our approach follows a probabilistic path: the proposed method quantifies qualitative relations as probability measures based 
on crowdsourced multiple observations contained in texts. Each observation is roughly quantified using a spatial feature vector comprising distance 
and orientation. Then, a greedy Expectation Maximization-based (EM) method is used to train a probability distribution. The latter represents the quantified 
spatial relationships under a probabilistic framework, i.e., it provides a set of random variables (spatial feature vector) that have certain probability density 
functions (PDFs) associated with them, for a specific spatial relation.

%\textbf{Contributions:} 
In this work, we employ probabilistic models to represent spatial relationships. To the best of our knowledge, this is the first work 
that combines qualitative and quantitative spatial information for spatial probabilistic inference. The novelty of our approach lies in the process of mapping textual crowdsourced and uncertain location 
observations to their probable locations based on probabilistic spatial relationship models. The traditional machine learning techniques that we employ have never been used to achieve
such a mapping. Moreover, this approach is one of the pivotal steps in developing automatic map-generation-from-text tools based on crowdsourced data. 

The outline of the remainder of this work is as follows. Section~\ref{sec:related_work} discusses related work. Section~\ref{sec:contribution} discusses the specific qualitative data involved and introduces the 
spatial feature vectors used for quantification, while Section~\ref{sec:statmodeling} introduces the tools necessary to derive quantification in the form of PDFs for the spatial 
relationships. Section~\ref{sec:experimentation} validates the proposed approach by means of “mapping” the obtained quantified relationships and using a similarity metric to assess the iterative quantification process. 
Finally, Section~\ref{sec:conclusions} presents conclusions and directions for future work.

\section{Related Work} 

% (fold)
\label{sec:related_work} 
%Work relevant to this research includes the extraction of semantic or especially spatial relations from natural language expressions, qualitative modeling of 
%spatial relations and its application to spatial databases, and, finally, quantitative modeling of spatial knowledge.

Work relevant to this paper includes qualitative modeling of spatial relations with application to spatial data management, 
and quantitative modeling of spatial knowledge.

\textbf{Qualitative:} The majority of works related to qualitative approaches for spatial information representation considers spatial relations. One popular spatial classification is constructed 
by topological relations (e.g., disjoint, overlap), direction relations (e.g., North, South), ordinal relations (e.g., inside, contain), and distance relations (e.g., far, near). 
The authors in \cite{egenhof2,egenhof1,egenhof3,Kainz} present formal methods for qualitative representation of spatial relationships based on mathematical theories of order. 
Their applicability on spatial database systems and some key-role technical concepts are coherently discussed in \cite{Guting, Papadias94theretrieval, Papadias95topologicalrelations}. %\cite{egenhof2,egenhof1,egenhof3,Kainz} 
%Subdivisions of land are represented as partially ordered sets, 
%named posets, a model that is general enough to answer spatial queries about inclusion and containment of spatial areas. After a brief introduction to the 
%basic concepts of posets and lattices, their applications to modelling spatial relations and operations for spatial regions in terms of containment and 
%overlay are presented. 
Qualitative representation of spatial knowledge is discussed in \cite{chorochron1, chorochron2, Papadias}. The authors identify the common concepts of the qualitative representation 
and processing of spatial knowledge. % underlying qualitative spatial knowledge representation. %\cite{chorochron1, chorochron2, Papadias}
They compare the representational properties of different systems and outline the computational tasks involved in relation-based spatial information 
processing. 
%They also describe symbolic spatial indexes, relation-based structures that combine several ideas in spatial knowledge representation. 

\textbf{Quantitative:} Recent research on quantitative representation of spatial knowledge has been conducted in relation to situational awareness systems, robotics, and image processing. 
Modelling uncertain spatial information for situational awareness systems is discussed in \cite{Kalashnikov1} and \cite{Kalashnikov2}. The authors  propose a bayesian probabilistic 
approach to model and represent uncertain event locations described by human reporters in the form of free text. They analyze several types of spatial queries of interest in situational 
awareness applications. Estimation of uncertain spatial relationships in robotics is addressed in \cite{roboticsquantrep}. The paper describes a representation of spatial information, 
called the stochastic map, and associated procedures for building it, reading information from it, and revising it incrementally as new information is obtained. The stochastic map contains 
the estimates of relationships among objects in the map, and their uncertainties, given all the available information. 
A probabilistic algorithm for the estimation of distributions over geographic locations is proposed in \cite{im2gps}. The authors use 
a data-driven scene matching approach in order to estimate geographic information based on images. 
In \cite{Vasudevan07abayesian} the authors attempt to create a hierarchical probabilistic concept-oriented 
representation of space, based on objects. Their approach is based on learning from exemplars, clustering and the use of Bayesian network classifiers. Such a conceptualization 
and representation can enable robots to be more cognizant of their surroundings. 
%Uncertain relations between imprecise regions are discussed in \cite{uncertaintoprels}. 
%The authors model the uncertainty in abstraction as well as the imprecision of measurement under a statistical framework. This allows introducing the uncertainty of observation 
%into qualitative spatial reasoning. 
Image similarity based on quantitative spatial 
relationship modeling is addressed in \cite{Wang}. The authors propose a novel method for the representation of relative spatial relations between objects in images, applied to 
multimedia database applications. 
%\cite{Kalogerakis} tries to achieve geolocation prediction under a machine learning framework by using Flickr photos. 
Finally, there has been some theoretic work on modeling spatial uncertainty using heuristics and fuzzy logic techniques. 
For example, in \cite{fuzzyquantrep}, a fuzzy decision tree algorithm is proposed to formalize spatial relations between linear objects.

%In this work, we employ probabilistic models to represent spatial relationships. The novelty of our approach lies in the process of mapping textual crowdsourced and uncertain location 
%observations to their probable locations based on probabilistic spatial relationship models. The traditional machine learning techniques that we employ have never been used to achieve
%such a mapping. This approach is one of the pivotal steps in developing automatic map-generation-from-text tools based on crowdsourced data. 
%%We have above summarized the existing body of research most related to our paper.Other concepts and techniques which are related to our work as well, will be discussed in this paper when the need arises.

% Discuss related work one paper at a time. For each paper briefly discuss its contribution and then state how your work differs from/improves over it.
% 
% The section concludes by saying overall how your approach improves over all existing ones.
% section related_work (end)
%!TEX root = paper.tex

%\section{Quantitative vs. Qualitative Data} 
\section{Spatial Features from Qualitative Data}
\label{sec:contribution}

The main contribution of this work is to model qualitative spatial information in a quantitative and probabilistic way. 
Our main data source will be narratives and this section will survey our approach for extracting qualitative data from texts. 
Moreover, to be able to quantify qualitative spatial data, we need to have a respective means for representing it. 
Here, we present the spatial feature vector that models spatial relationships based on distance and orientation measures.  

\subsection{Dataset}
\label{subsec:data} 
Crowdsourced narratives are likely to contain spatial information. The more relevant 
the text is to ``space'', the more data it will contain. Our specific case considers 
travel blogs as a rich potential data source. This assumption is based on the 
intuition that people tend to describe their experiences in relation to their 
trips and places they have visited. This behavior results in ``spatial'' narratives. 

To obtain such data, we used classical web crawling techniques as presented in \cite{Drymonas} and we compiled a database\footnote{Available upon request} consisting 
of 120K user generated texts obtained from travel blogs\footnote{TravelBlog, TravelJournal, TravelPod}.
%such as \textit{TravelBlog, TravelJournal and TravelPod}. 
%All these texts are narratives generated from users who describe places they have visited all over the world. 

\subsection{Spatial Relations}
\label{subsec:data} 

Obtaining qualitative spatial data from text involves the detection $(i)$ of spatial objects, i.e., Points-of-Interest (POIs) 
or toponyms and $(ii)$ of spatial relationships between those POIs.
The employed approach involves geoparsing, i.e., the detection of candidate phrases, and geocoding, i.e., 
linking the phrase/toponym to actual coordinate information. 
Using GATE's \cite{Gate} text processing and semantic analysis components 
in combination with the algorithm presented in \cite{Drymonas}, we managed to extract 120k POIs from the text corpus.  
For the geocoding of the POIs, we rely on the open-source module GeoGoogle\footnote{http://geo-google.sourceforge.net}, 
a Java API utilizing the geocoding service that is part of the Google Maps API. This procedure associates (whenever possible) geographic coordinates with POIs that have been identified in the travel blog data. 
The final result of this stage is an index that contains the geographic coordinate information plus text information of all POIs, i.e., document, paragraph, sentence and word distance information.
%Figure~\ref{fig:poismap} illustrates the spatial distribution of the Points-of-interest (POIs) that were found in the texts. 

The following excerpts are texts that contain relevant POIs and respective spatial relationship data. 
\begin{itemize}
	\item ``...and then went out for tea at a lovely Italian restaurant in \textit{Soho} \textbf{near} \textit{Covent Garden}'' 
%	\item ``The last thing we looked at was the outside of \textit{St. Pancras train station} and  \textit{Kings Cross station},which is \textbf{neighboring it}.'' 
%	\item ``Our little apartment is located in \textit{Bayswater} \textbf{next to} \textit{Notting Hill}.''
	\item ``\textit{Tate Modern} is a big modern art gallery, \textbf{on} \textit{South Bank}, amazing building, has some great stuff in it as well.''
%	\item  ``\textit{Piccadilly Circus} is just \textbf{around the corner} from the posh shopping area of \textit{Regent Street} and \textit{Oxford Street}.''
\end{itemize}

These examples confirm our initial hypothesis for the existence of spatial knowledge in user generated content. As expected, %n Figure~\ref{fig:poismap} one can easily deduce that
we observe that POIs are more dense in urban places. %In the experimental section, we will explain that this is the reason why we consider data from such regions as more informative. 

Having identified and geocoded the spatial objects, the next step is the localization of qualitative spatial relationships. 
This would ideally require efficient natural language processing (NLP) tools to automatically extract and map phrases to 
spatial relations linking POIs as contained in texts. 
%Automatic extraction of spatial knowledge from texts is a very hard NLP problem 
%and unfortunately none of the contributions on semantic and spatial relation extraction from text are either adequate and/or available to accomplish this task. 
Kernel methods for semantic relation extraction between entities in texts are developed in \cite{bunescu2006subsequence} and \cite{Zelenko}. %\cite{Zelenko, bunescu2006subsequence} 
In \cite{extrir} and \cite{svmextr}, Support Vector Machine (SVM) approaches are used to extract spatial relations for spatial reasoning. %\cite{extrir, svmextr}
%A SVM based extraction of spatial relations from text is addressed in \cite{svmextr}, too.
%where the authors investigate the extraction of spatial relations based only on SVM models.
%which can implement the recognition of spatial expressions and their classification synchronously. 
%For the SVM model, a set of feature vectors are specified, such as lexical tokens, 
%spatial terms, syntactic structures and geographical feature types of place names, and a multi-label classifier is presented to solve the multi-classification problem.
In \cite{spatialrolelabel}, the authors report on a novel task of spatial role labeling in text, 
based on machine learning methods to extract spatial roles and their relations. 
%More specifically, they use a probabilistic approach by training Conditional Random Fields (CRFs),
%which is a special case of Markov Random Fields, in order to achieve spatial information extraction. 
Finally, extraction of semantic relations from texts using dependency grammar patterns is addressed in \cite{ReVerb} and \cite{wanderlust09}. 
%The authors present Wanderlust, an algorithm that automatically extracts semantic relations 
%from natural language text by using deep linguistic patterns. 
Overall, while several of these techniques would be useful for spatial relationship extraction from texts, 
none either performed in a satisfying way or were available to us.
%which are defined over a dependency grammar of sentences. 
% Due to its linguistic nature, the method performs in an unsupervised fashion 
% and is not restricted to any specific type of semantic relation. The applicability of the proposed approach is examined in a case study, in which it is put to the task of generating 
% a semantic wiki from the English Wikipedia corpus. The algorithm is based on the hypothesis that 46 universally valid grammatical patterns can be used to extract spatial (semantic) 
% relations from plain text with a precision of 80\%. 

%
For the scope of this work, which lies on the area of probabilistic modelling of qualitative data but not on the extraction of qualitative spatial knowledge from text, 
we overcome this problem by using human annotation in combination with filtering of the input dataset. We 
restrict the data to be considered $(i)$ to the geographic area of London and further reduce it by $(ii)$ only considering sentences 
that include at least two POIs (which are needed to express a spatial relationship).
This manageable dataset (sentences) is then annotated by humans to extract spatial relationships. 
Human annotation results into tuples of the form shown in Table~\ref{table:tuples}. Here $\mathcal{P} = \{P_1, \ldots, P_m\}$ represents 
the set of spatial objects participating in binary spatial relationships $\mathcal{R} = \{R_1, \ldots, R_n\}$ with $i,j \leq m$ and $k \leq n$.

%We consider that $P_u \in \mathcal{U}$ and $P_k \in \mathcal{K}$ as described in Section~\ref{sec:introduction}. Both $u$ and $k$ are bounded between $1$ and $m-1$ because 
%a given set of objects should contain at least one but less than $m$ known objects, otherwise it is not feasible to predict possible locations of unknown objects,  
%and at least one but less than $m$ unknown objects.   

% and we declare a point of interest that is participates in a relation with another reference point of interest $p_j$ using a spatial relationship $r_k$ out of a total of $n$ relationships we consider. 
%Here $n$ denotes the total number of spatial relationships that appear in the travelblog dataset. 
%For each row $p_i \neq p_j$. Additionally, there are no pair ($p_i,p_j$) duplicates for each spatial relationship $r_k$, where $1 \leq k \leq n$. However, there may exist but some pairs could be connected 
%with more that one spatial relationships.     

\begin{table}[htbp]
\begin{center}
    \begin{tabular}{|l|l|l|}
        \hline
  $\mathcal{P}$ & $\mathcal{R}$ & $\mathcal{P}$ \\ \hline
  $P_{1}$ & $R_{1}$ & $P_{2}$ \\ 
  $P_{3}$ & $R_{1}$ & $P_{4}$ \\
  $P_{3}$ & $R_{1}$ & $P_{5}$ \\
   \vdots & \vdots  &  \vdots \\
  $P_{i}$ & $R_{k}$ & $P_{j}$ \\
\hline
    \end{tabular}
\end{center}
\caption{Dataset denoting (geocoded) spatial objects and spatial relations.}
\label{table:tuples}
\end{table}

\subsection{Spatial Features}
\label{subsection:spatfeats}

Statistical models are often used to represent observations in terms of random variables. These models can then be used for estimation, 
description, and prediction based on basic probability theory. 
In our approach, we model a spatial relation between two POIs $P_k \in \mathcal{K}$ and $P_u \in \mathcal{U}$ 
($k$ declares known and $u$ declares unknown as described in Section~\ref{sec:introduction}) in terms of \emph{distance} and \emph{orientation}. 
We consider a labeled \emph{spatial feature vector} as two random variables that model spatial relations in a probabilistic way. 
Assuming a projected (Cartesian) coordinate system, the distance is computed as the Euclidean metric between the two respective coordinates. 
The orientation is established as the counterclockwise rotation of the x-axis, centered at $P_k$, to the unknown point $P_u$. 

Several instances of a spatial relation are used to create a dataset which will be used to train a probabilistic  %(cf. Table~\ref{table:tuples}),
model for each spatial relation. Under a mathematical formalization, let us consider that for each instance of each 
relation we create a two-dimensional spatial feature vector $X = (X_d, X_o)^{\intercal}$ where $X_d$ denotes the distance and $X_o$ denotes the orientation between $P_k$ and $P_u$.
We end up with a set of two-dimensional feature vectors $\mathcal{X}=\{X_1, X_2, \dots, X_n\}$ for each spatial relation where the $i$-th vector of 
each set has the form $X_i = (X_{di}, X_{oi})^{\intercal}$. An example of the feature extraction procedure is illustrated 
in Figure~\ref{fig:featext}, where four instances of spatial relation \textit{Near} are used in order to create the respective 
set of spatial feature vectors $\mathcal{X}_{near}=\{[X_{d1}, X_{o1}]^{\intercal}, [X_{d2}, X_{o2}]^{\intercal}, [X_{d3}, X_{o3}]^{\intercal}, [X_{d4}, X_{o4}]^{\intercal}\}$. In this scenario, 
$\mathcal{K}=\{A, D, E, G\}$ and $\mathcal{U}=\{B, C, F, H\}$.% and $\mathcal{R}=\{'Near'\}$.
%set as shown in Table~\ref{table:nrcase}. 
%
%A feature vector set of this form is calculated for each spatial relation allowing us to create a dataset that can be used to train a 
%probabilistic model for each one of them.

% \begin{figure}[here]
% 	\includegraphics[width=2.8in,height=2.2in]{drawings/NearCaseFeats.pdf} \caption{Distance and Orientation Feature Extraction} \label{fig:featext} 
% \end{figure}

\tikzset{
  big arrow/.style={
    decoration={markings,mark=at position 1 with {\arrow[scale=1.5,#1]{>}}},
    postaction={decorate},
    shorten >=0.2pt},
  big arrow/.default=black}

\begin{figure}
\begin{center}

\scalefont{0.5}
\begin{tikzpicture}[scale=1.0]
  % Draw axes
    \draw [<->,thick] (0,7) node (yaxis) [above] {$y$}
        |- (10,0) node (xaxis) [right] {$x$};
    \coordinate[label=200:$A$]  (a) at (1.0,1.0);
    \coordinate[label=45:$B$]   (b) at (2.0,2.2);
    \coordinate[label=290:$C$]  (c) at (1.5,4.2);
    \coordinate[label=290:$D$]  (d) at (3.0,5.0);
    \coordinate   (e) at (6.5,5.5);
    \coordinate[label=0:$F$]   (f) at (7.5,4.8);
    \coordinate[label=290:$G$]  (g) at (8.0,1.0);
    \coordinate[label=45:$H$]  (h) at (6.5,3.3);
    
    \coordinate[label=290:$E$] (k) at (6.5,6.0);
    
    \fill[black] (a) circle (1.2pt);
    \fill[black] (b) circle (1.2pt);
    \fill[black] (c) circle (1.2pt);
    \fill[black] (d) circle (1.2pt);
    \fill[black] (e) circle (1.2pt);
    \fill[black] (f) circle (1.2pt);
    \fill[black] (g) circle (1.2pt);
    \fill[black] (h) circle (1.2pt);
    
    \draw[black, big arrow] (a)--(b);
    \draw[black, big arrow] (d)--(c);
    \draw[black, big arrow] (e)--(f);
    \draw[black, big arrow] (g)--(h);
    
    \draw [dashed] (0.5,1)--(3,1);
    \draw [dashed] (1,2.5)--(1,0.5);
    
    \draw [dashed] (3,6)--(3,4.5);
    \draw [dashed] (2.5,5)--(4.5,5);
    
    \draw [dashed] (6.0,5.5)--(8,5.5);
    \draw [dashed] (6.5,5)--(6.5,6.5);
    
    \draw [dashed] (8,0.5)--(8,2.5);
    \draw [dashed] (7.5,1.0)--(9.7,1.0);
    
    \draw [black!50!black,thick](a) +(0:.3cm) arc (0:37:.4cm);
    \draw [color=black](a)+(15:0.65) node[rotate=0] {$X_{o1}$};
    
    \draw [black!50!black,thick](d) +(0:.2cm) arc (0:210:.2cm);
    \draw [color=black](d)+(135:0.55) node[rotate=0] {$X_{o2}$};
    
    \draw [black!50!black,thick](e) +(0:.2cm) arc (0:330:.2cm);
    \draw [color=black](e)+(145:0.55) node[rotate=0] {$X_{o3}$};
    
    \draw [black!50!black,thick](g) +(0:.2cm) arc (0:120:.2cm);
    \draw [color=black](g)+(35:0.55) node[rotate=0] {$X_{o4}$};
    
    \draw [decorate,decoration={brace,amplitude=5pt},xshift=-4pt,yshift=0pt]
    (a)--(b) node [black,midway,xshift=-0.15cm,yshift=+0.3cm] {$X_{d1}$};
    %{\footnotesize {$X_{d1}$};
    
    \draw [decorate,decoration={brace,amplitude=5pt},xshift=-4pt,yshift=0pt]
    (d)--(c) node [black,midway,xshift=+0.32cm,yshift=-0.32cm] {$X_{d2}$};
    %{\footnotesize $X_{d2}$};
    
    \draw [decorate,decoration={brace,amplitude=5pt},xshift=-4pt,yshift=0pt]
    (f)--(e) node [black,midway,xshift=+0.0cm,yshift=-0.33cm] {$X_{d3}$};
    %{\footnotesize $X_{d3}$};
    
    \draw [decorate,decoration={brace,amplitude=5pt},xshift=-4pt,yshift=0pt]
    (g)--(h) node [black,midway,xshift=-0.3cm,yshift=-0.2cm] {$X_{d4}$};
    %{\footnotesize $X_{d4}$};
\end{tikzpicture}
\caption{Distance and orientation feature extraction procedure. In this case B is near A, C is near D, F is near E and H is near G.} \label{fig:featext}
\end{center}
\end{figure}
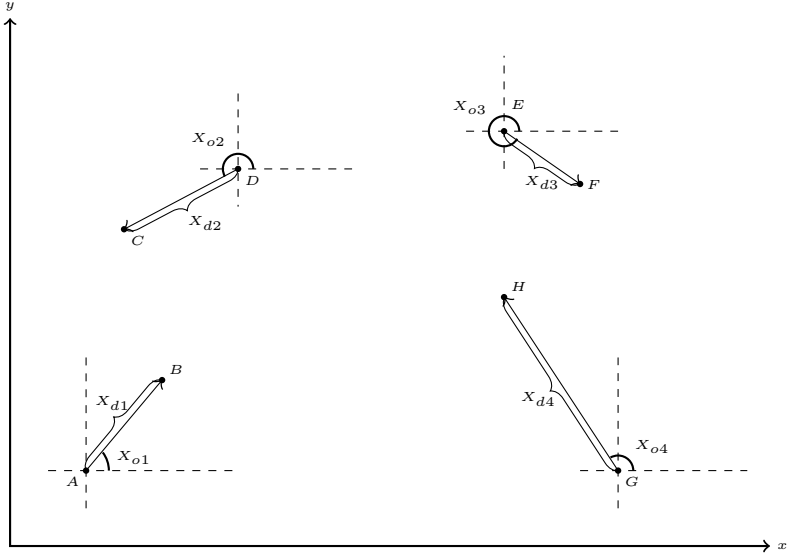

% \begin{table}[h]
% \begin{center}
%    \begin{tabular}{|l|l|l|l|l|}
%         \hline
% 		  & Near   & Near   & Near   & Near   \\ \hline
%         Distance         & $X_{s 1}$ & $X_{s 2}$ & $X_{s 3}$ & $X_{s 4}$ \\ 
%         Orientation      & $X_{\alpha 1}$ & $X_{\alpha 2}$ & $X_{\alpha 3}$ & $X_{\alpha 4}$ \\
%         \hline
%     \end{tabular}
% \end{center}
% \caption{Feature vector set for spatial relation \textit{Near} generated from the instances illustrated in Figure~\ref{fig:featext}}
% \label{table:nrcase}
% \end{table}

%!TEX root = paper.tex
\section{Spatial Relation Modeling} 
\label{sec:statmodeling}

In this section we discuss the methods and algorithms we used to train probabilistic models that 
can efficiently represent spatial relationships based on our dataset. 
More specifically, we start by describing the approach we use to populate our dataset by using Kernel Density Estimation (KDE)
which is a state-of-the-art method for the estimation of a multi-dimensional probability density function.  
We continue by analyzing the Gaussian Mixture Model (GMM), which is the probabilistic model we employ for the quantitative 
representation of spatial relations and we outline a greedy learning algorithm for parameter estimation of the GMM. 
Finally, we discuss Kullback-Leibler (KL) divergence,  %a probability density distance metric 
which is utilized to assess the similarity between GMMs, i.e., 
comparing GMM estimation stages of the same spatial relation, 
but also to compare different relations.

\subsection{Populating a Spatial Feature Dataset} 
\label{subsection:kde}

The collected data includes 120K texts from travel blogs; however, with a focus on a specific geographic area (London), the dataset does not include enough spatial 
relationship instances to train a two-dimensional probabilistic model.
%As we mentioned in section ~\ref{sec:contribution}, we managed to collect $120000$ user generated texts from travel blogs. Parsing and semantic analysis of these texts resulted in $120000$ points of interest. 

To obtain more data we use KDE \cite{applsmoothtech}. %available for training a two-dimensional probabilistic model 
In our scenario, KDE techniques provide new density samples based on a small amount of ground-truth data. These estimates are 
then used in order to generate additional spatial feature vector data (semi-synthetic) to train probabilistic models (GMMs).  

Relating KDE to our problem, let $X = (X_d, X_o)$ follow a two-dimensional true density $f$ defined over $ \mathbb{R}^{2}$. 
%As mentioned in section ~\ref{sec:contribution}, $X_1$ which represents distance and $X_2$ is a random variable which represents orientation. 
Let $\mathcal{X}=\{X_1,$\dots$,X_n\}$ be an independent random sample set (initial spatial feature vector set in our case) 
drawn from $f$. The general form of the kernel density estimation function of $f$ is:

\begin{equation}
	\hat{f}_{H}(x;H) = \frac{1}{n} \sum_{i=1}^{n} K_{H}(x-X_i) 
\end{equation} where $x=(x_1,x_2)^{\intercal}$ is a generic vector that depends on the Kernel used, e.g. Gaussian, Epanechnikov, Cosine etc., $X_i = (X_{d i},X_{o i})^{\intercal}$ with $1 \leq i \leq n$,   
$K_{H}(x) = |H|^{-\frac{1}{2}} K(H^{-\frac{1}{2}} x)$, and $n$ denotes the number of instances of each spatial relation. 
In our case, $K_{H}(x)$ is a Gaussian bivariate kernel function, and $H$ is a symmetric positive definite $2\times 2$ diagonal 
matrix (bandwidth matrix).

The performance of a kernel density estimator is primarily determined by the choice of bandwidth, which controls the degree of smoothing, 
and secondarily by the choice of the kernel function, which in our case is Gaussian.
A large body of literature \cite{applsmoothtech,bandwidthselMV} exists on bandwidth selection for univariate %,Turlach_bandwidthselection  %\cite{applsmoothtech,Duong,bandwidthselMV}
and multivariate kernel density estimation. In this contribution, we follow a simple case scenario as described in \cite{applsmoothtech}. 
With data observed from a bivariate normal density, the diagonal bandwidth matrix, denoted by 

\begin{equation}
 H = \begin{pmatrix}
  h_1 & 0\\
  0 & h_2
  \end{pmatrix}
\end{equation} can be well approximated by $h_b = \sigma_b \left(\frac{4}{(d+2)n} \right)^{\frac{1}{d+4}}$ 
for $b \in \lbrace 1,2\rbrace$, where $\sigma_b$ is the standard deviation of the $i$-th variate and $d$ denotes the problem's dimensionality. %(in our case $d=2$) 
This method is often used when no other practical bandwidth selection scheme is available, despite the fact that most interesting data are non-Gaussian. 

%In this work, we generated $200$ samples for each extracted spatial relation. 

%
% Data of such a process is visualized in Figure~\ref{fig:kde}. Figure~\ref{fig:kde}(a) illustrates the initial dataset for a spatial relationship 
% in a two-dimensional space (distance [km] and orientation [degrees] and the $x$ and $y$-axis, respectively). Figure~\ref{fig:kde}(b) shows the 
% Gaussian kernel density estimate of the probability density function (PDF) of the dataset shown in Figure~\ref{fig:kde}(a). 
% Finally, Figure~\ref{fig:kde}(c) shows the generated samples using the estimated PDF. 

\begin{figure}[h] 
\begin{center}
\includegraphics[width=4.0in,height=3.8in]{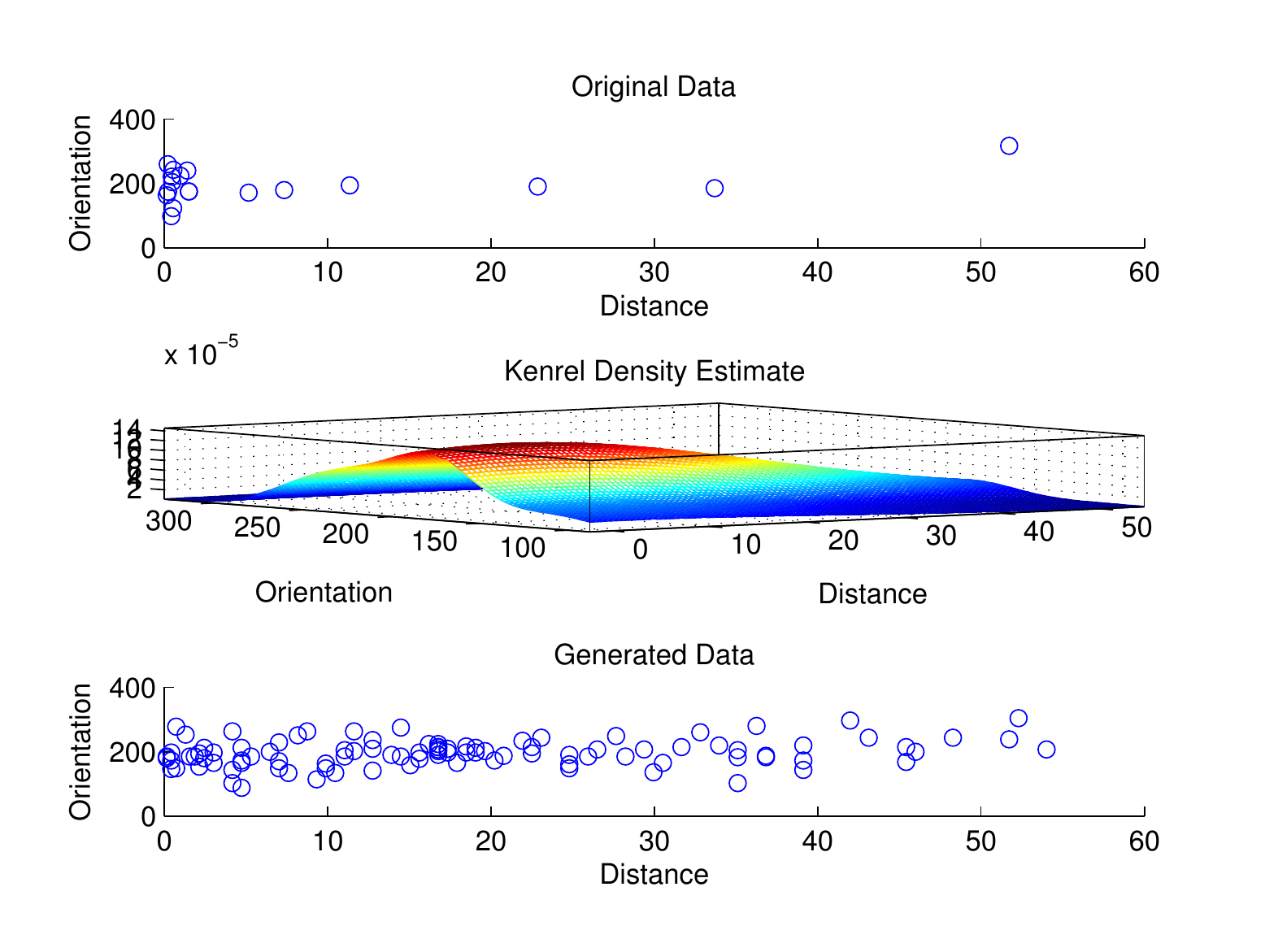} 
\caption{Spatial feature dataset population: $(i)$ initial dataset $(ii)$ estimated probability density function using KDE, and $(iii)$ generated dataset.} 
\label{fig:kde} 
\end{center}
\end{figure}

Data of such a process is visualized in Figure~\ref{fig:kde} which illustrates the initial dataset for a spatial relationship 
in a two-dimensional space (distance [km] and orientation [degrees] as the $x$ and $y$-axis respectively), the 
Gaussian kernel density estimate of the probability density function (PDF) of the initial dataset and 
the generated samples using the estimated PDF, respectively.

Finally, as we explain in Section~\ref{sec:experimentation}, we underline that we use 
the generated data only for training probabilistic models and not for testing, 
since the generated data exhibits a considerable bias.

% \begin{figure*}[htp]
% \centering
% \subfigure[]{\includegraphics[width=8.0cm,height=8.0cm]{drawings/datagen.pdf} \label{fig:label_a}} \quad
% \subfigure[]{\includegraphics[width=8.3cm,height=7.5cm]{drawings/em.pdf} \label{fig:label_b}}
% \caption{(a) Spatial feature dataset population. Initial dataset, probability density function using KDE and populated dataset, respectively. (b) Converged max 3-Component GMM}
% \label{fig:label} 
% \end{figure*}

\subsection{Quantifying Qualitative Relations}
\label{subsection:gmmem}

The essential step in quantifying qualitative data is the mapping of the generated data to PDFs.
%how is that different from KDEs in the previous step?
Here we have to decide what kind of probabilistic model we desire to train. Using Gaussian kernel density estimation to populate our dataset, 
we naturally opted for Gaussian Mixture Models (GMMs).
GMMs have been extensively used in many classification and general machine learning problems (cf. \cite{Bishop,Duda}). 
They are very well known for $(i)$ their formality, as they build on the formal probability theory, $(ii)$ their practicality, as they have been implemented 
several times in practice, $(iii)$ their generality, as they are capable of handling many different types of uncertainty, and $(iv)$ their effectiveness 
because existing solutions that employ them known to be effective and scalable.
 
Generally speaking, a GMM is a weighted sum of $M$ component Gaussian densities as 

\begin{equation} \label{eq:gmm}
  p(x|\lambda) = \sum_{i=1}^{M} w_{i} g(x;\mu_{i},\Sigma_{i})
\end{equation} where $x$ is a d-dimensional data vector (i.e., features - in our case $d=2$), $w_i$ , $1 \leq i \leq M$, are the mixture weights, and $g(x|\mu_i, \Sigma_i )$ 
is a Gaussian density function $\forall i$, with mean vector $\mu_i \in \mathbb{R}^{d} $ and covariance matrix $\Sigma_i \in \mathbb{R}^{d\times d}$  such that
%The mixture weights satisfy the constraint that  $\sum_{i=1}^{M} w_i = 1$ with $w_{i}\geq0$.

\begin{equation} \label{eq:gaussdens}
\begin{split}
	& g(x;\mu_i, \Sigma_i) = \\
	 & (2\pi)^{-\frac{d}{2}} \det{(\Sigma_i)}^{-\frac{1}{2}} \exp{(-\frac{1}{2}(x-\mu_i)^{\intercal}\Sigma_i^{-1}(x-\mu_i))} 
\end{split}
	\end{equation}

with mean vector $\mu_i \in \mathbb{R}^{d} $ and covariance matrix $\Sigma_i \in \mathbb{R}^{d\times d}$. The mixture weights satisfy the constraint that 

\begin{equation}
 \sum_{i=1}^{M} w_i = 1 
\end{equation}

with $w_{i}\geq0$.

The complete GMM is parameterized by the mean vectors, the covariance matrices and mixture weights
$\forall i$. These parameters are collectively represented in Equation~\ref{eq:gmm}, 
by the notation $\lambda = \{w_i, \mu_i, \Sigma_i\}$ with $i= 1, \dots, M$. 
In our setting, each spatial relation is modeled by a 2-dimensional GMM trained with each relation's spatial feature
vectors, which are created as detailed in Sections~\ref{subsection:spatfeats} and \ref{subsection:kde}.

% This is beacuse in this work we are interested in prooving that distance and orientation features are 
% informative enough to model spatial relationships between objects in a real e Cartesian framework. Continuing,
%  we briefly describe the mathematical nature of Gaussian Mixture Models (GMMs) are their use in classical machine 
% learning and probability theory. Finally, we briefly analyze a state of the art greedy Expectation Maximization (EM) 
% approach as presented in \cite{Verbeek}, which is used in order to efficiently train GMMs and precisely decide about 
% the number of components that should be used for each spatial relation's trained GMM. As we mentioned before, 
% in this part of this contribution we briefly describe the nature of Gaussian mixture 
% models as presented in \cite{Duda, Bishop} and the usedfullness of the Expectation Maximization algortith for 
% model parameter estimation as presented in \cite{Dempster}. 

For the parameter estimation of Gaussian component of each GMM, we use Expectation Maximization (EM) (cf. \cite{Dempster}). 
EM enables us to update the parameters of a given M-component mixture with respect to a feature vector 
set (generated spatial feature vector set in our case) $\mathcal{X} = \{X_1 , \dots, X_n \}$ with $1 \leq j \leq n$ and all $X_j \in \mathbb{R}^{d}$, such that 
the \emph{log-likelihood} $\mathcal{L}$ of $\mathcal{X}$ calculated using Equation~\ref{eq:loglike} increases with each re-etimation step. 
This means that we keep re-estimating model parameters until the log-likelihood $\mathcal{L}$ or the parameters converge. 
 
\begin{equation}
	\label{eq:loglike} 
	\mathcal{L} = \sum_{j=1}^{n} \log(p(X_j|\lambda)) 
\end{equation}

The updates for the parameters of a GMM can be accomplished by iterative application of the following equations 
for all components $i\in \{1, . . . , M\}$

\begin{equation}
	P(i|X_j) = \frac{w_i g(X_j;\lambda_i)}{p(X_j|\lambda)} 
\end{equation}

\begin{equation}
	w_i = \sum_{j=1}^{n} \frac{P(i|X_j)}{n} 
\end{equation}

\begin{equation}
	\mu_i = \sum_{j=1}^{n} \frac{P(i|X_j)X_j}{n w_i} 
\end{equation}

\begin{equation}
	\Sigma_{i} = \sum_{j=1}^{n} \frac{P(i|X_j)(X_j - \mu_i)(X_j-\mu_i)^{\intercal}}{n w_{i}} 
\end{equation}

The EM algorithm is not guaranteed to lead us to the solution yielding maximum log-likelihood on $\mathcal{X}$ among all maxima 
of the log-likelihood. Nevertheless, using the EM algorithm, if we are ``close'' to the global optimum (maximum) 
of the parameter space, then it is very likely we can obtain the globally optimal solution.

\subsection{Model Optimization}
\label{subsection:modopt}

A main issue in probabilistic modeling with GMMs is that a predefined number of components per Gaussian mixture 
is neither a dynamic nor an efficient and robust approach. The optimal number of Gaussian components should be decided based on each dataset.  
Hence, in this section we employ a greedy learning approach to dynamically estimate the number of components in a GMM. (cf. \cite{Verbeek}). 
Typically a GMM is trained by starting with a random configuration of all components and improve upon this configuration with the EM algorithm. 
This greedy approach tries to build the mixture component in a more efficient way by starting from an one-component GMM, whose parameters are 
trivially computed by using EM (cf. Section~\ref{subsection:gmmem}), and then employing the following two steps until a stop criterion is met. 

\begin{enumerate}
\item Insert a new component in the mixture 
\item Apply EM until the log-likelihood $\mathcal{L}$ or the parameters of the GMM converge (cf. Section~\ref{subsection:gmmem})
\end{enumerate}

The stop criterion can either be a maximum pre-specified number of components, or it can be any other model selection criterion like 
Minimum Description Length \cite{Grunwald} or Bayesian Information Criterion (BIC) \cite{citeulike}.
In our case the algorithm stops if the maximum number of components is reached, 
or if the log-likelihood $\mathcal{L}$ after introducing a new component is lower than that of the previous model. For a more formal description let us 
consider a feature vector set (generated spatial feature vector set in our case)  $\mathcal{X}$ under an M-component mixture $p^{M}(\mathcal{X}|\lambda)$. 
The greedy learning algorithm can be summarized in the following five steps:

\begin{enumerate}
\item Compute the one-component mixture $p^{1}(\mathcal{X}|\lambda)$  that yields maximum log-likelihood using the (EM) algorithm. 
\item Find the optimal new component $g(\mathcal{X};\lambda^{\ast})$ and the corresponding mixing weight $w^{\ast}$.
\item Set $p^{M+1}(\mathcal{X}|\lambda) = (1- w^{\ast})p^{M}(\mathcal{X}|\lambda) + w^{\ast} g(\mathcal{X};\lambda^{\ast})$ and $M = M+1$.
	
	%(1 − {\alpha}^{\ast})f_k(x) + \alpha^{\ast} \phi(x;\theta^{\ast})$ and $k := k + 1$
	
\item Update new model parameters using EM algorithm. 
\item Terminate if ($i$) log-likelihood $\mathcal{L}$ of GMM starts to decrease, or ($ii$) max number of components is reached; else go to step 2. 
\end{enumerate}

The crucial step of the algorithm is the component insertion in Step 2. 
Several approaches exist here. One is to consider a number of candidates equal to the number of feature vectors but this would be rather 
expensive. The approach followed in this work is to pick an optimal number of candidate components as discussed in \cite{Verbeek}. 

\begin{figure}[here] 
\begin{center}
\includegraphics[width=3.7in,height=3.1in]{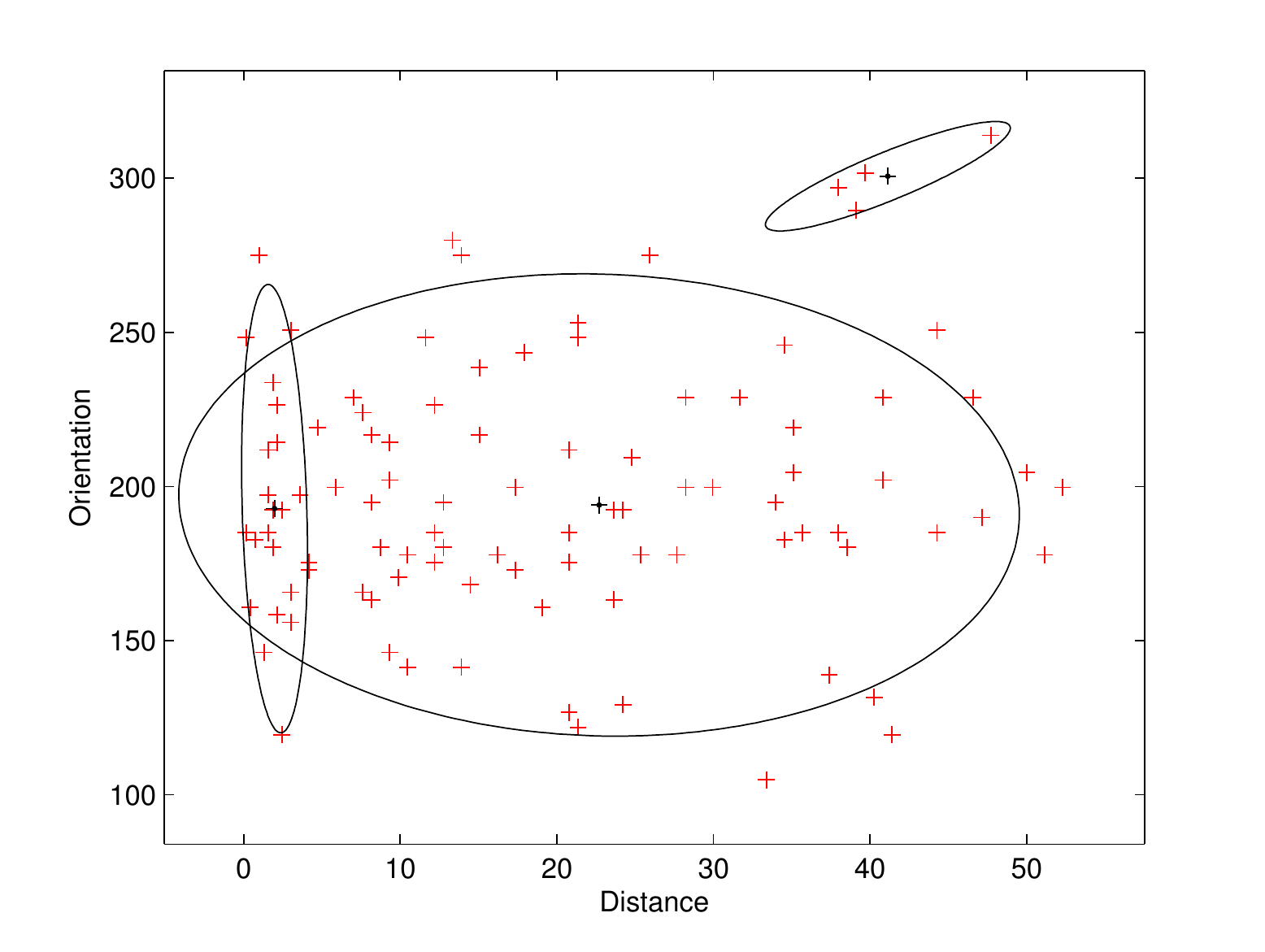} 
\caption{Converged max 3-component GMM.} 
\label{fig:gmm} 
\end{center}
\end{figure}

Figure~\ref{fig:gmm} illustrates a converged 3-component GMM.  In this case, the maximum number of Gaussian components was used as a stop criterion. 
Distance and orientation are used as uncorrelated random variables, which means that all Gaussian components in each GMM have diagonal covariance 
matrices. The $x$ and the $y$-axes represent raw (not normalized) distance and orientation information in kilometers and degrees, respectively.

\subsection{Similarity Between Quantitative Spatial Relationships} 
\label{subsection:kl}

In many probabilistic classification problems several metrics have been proposed to compute a distance measurement between different classes as a means to compare them.
Measuring distance between converged PDFs which model different classes (spatial relationships in our case) is a measure of similarity between them. 
In our contribution, we use Kullback-Leibler (KL) divergence \cite{Kullback} as such a distance metric. 

There are two main reasons for checking similarity between quantified spatial relations.
Firstly, we want to observe the changes for each GMM as we increase the maximum number of Gaussian components during the training procedure. 
Secondly, we use KL divergence to measure the similarity between spatial relationships that tend to follow similar patterns, e.g., \emph{Near} \& \emph{NextTo}, \emph{In} \& \emph{On}. 
%
%The approach is used in the experiments of Section~\ref{sec:experimentation}.
   
KL divergence is a similarity measure between two probability distributions. 
So, let $\mathcal{F}_1(x)$ and $\mathcal{F}_2(x)$ be two probability distributions (GMMs in our case). By definition, the KL distance $\mathcal{D}(\mathcal{F}_1(x)||\mathcal{F}_2(x))$ 
between $\mathcal{F}_1(x)$ and $\mathcal{F}_2(x)$ is given as follows.

\begin{equation}
	\mathcal{D}(\mathcal{F}_1(x)||\mathcal{F}_2(x)) = \int \mathcal{F}_1(x) \log{
	\begin{Bmatrix}
		\frac{\mathcal{F}_1(x)}{\mathcal{F}_2(x)} 
	\end{Bmatrix}
	} dx 
\end{equation}

The KL divergence is always nonnegative and it is zero only when the two distributions are identical. 
Additionally KL divergence is not symmetric, i.e., $\mathcal{D}(\mathcal{F}_1(x)||\mathcal{F}_2(x)) \neq \mathcal{D}(\mathcal{F}_2(x)||\mathcal{F}_1(x))$.

It is common to encounter the symmetric version of the KL divergence between $\mathcal{F}_1(x)$ and $\mathcal{F}_2(x))$ as
\begin{equation}
	\mathcal{D}_{sym}(\mathcal{F}_1(x)||\mathcal{F}_2(x)) = \frac{\mathcal{D}(\mathcal{F}_1(x)||\mathcal{F}_2(x)) + \mathcal{D}(\mathcal{F}_2(x)||\mathcal{F}_1(x))}{2} 
\end{equation}

In this work, we use the symmetric KL divergence in order to measure the similarity between GMMs.  
%!TEX root = paper.tex
\section{Experimentation} 
\label{sec:experimentation}

The scope of this section is to assess the quantitative representation of qualitative geospatial data by means of probability distributions (GMMs). 
For this purpose, we investigate a set of spatial relationships for a specific geographic area (London).
%Here, we conduct two sets of experiments. 
%
In terms of experiments, we compute probabilistic representations of spatial relationships by considering distance and orientation as dependent but, 
uncorrelated features (case one) and as correlated features (case two). 
%
%In practice, in the first case scenario we train GMMs with diagonal covariance matrices while in the second scenario we train GMMs with full covariance matrices (cf. Section~\ref{subsection:modopt}). 

We visualize the results of the trained models and compare them to check if they intuitively perform well, e.g.,  
they return visually reasonable results. 
%against models whose parameters where estimated optimally by maximizing the log-likelihoods of the data set during the training step. 
%
In addition, we measure the KL divergence for spatial relationships between a baseline one-component model and the maximum number of Gaussian components model. 
Finally, based on visualization and KL divergence, we assess the informativeness and efficiency of distance and orientation features for quantitative modeling 
of spatial relations and observe how much different spatial relations may behave in a similar way. 

% This section empirically verifies your approach. Here you have to show that your solution performs well.
% How do we do this? We compare our solution to the ``best'' existing solution using a specific set of parameters such as running time, number of I/Os, quality of result, etc.
% In the introduction to this section you clarify all those points by stating something like the following.
% The objective of the performance study is to evaluate the three map-matching algorithms in terms of (i) their running times and (ii) the quality of their respective matching results.

\subsection{Experimental Setup} 
\label{ssub:experimental_setup}

The choice of an appropriate dataset is crucial in our experimentation. As mentioned is Section~\ref{subsec:data}, the density of POIs is very high in urban regions. %Figure~\ref{fig:poismap} clearly shows regions on the map where the 
We decided to use data from such a dense region to find meaningful as well as consistent spatial relationships. We retrieved data for a bounding box that contains the greater 
area of London, UK. In this preprocessing step, we parsed our travel blog data (120k texts) set and retrieved sentences containing at least two POIs and whose coordinates are 
within the bounding box of Latitude $[51\degree,52\degree]$ and longitude $[-1\degree,1\degree]$. This resulted in 12k sentences. Using human annotation, we extracted instances 
of the eight most frequent spatial relations including \emph{North, South, East, West, Near, In, On, NextTo}.
This means also that in our travel blog dataset, people tend to use a mixture of directional, topology and vague metrical relations in order to describe POI locations. 
From this data, distance and orientation features where extracted as described in Section~\ref{sec:contribution}. 

%As in many research areas that are based on crowdsourcing, our basic problem here is that we need to collect a rather large amount of data overall to obtain a meaningful amount of useful data. 
Given that only a small percentage of the collected data contains spatial relationship information, here to obtain a meaningful amount of useful data, we need to overall collect a large volume of texts.  
For example, considering the London case, approximately 10\% of the 12000 sentences contained clear instances of %we managed to extract only 300 instances of 
spatial relationships. For the specific approach, this would have not been enough to train and test probabilistic models. 
We use KDE to create a semi-synthetic dataset based on the collected data (cf. Section~\ref{subsection:kde}). More specifically, 
we use KDE to estimate each spatial relationship's density function. 
This estimate is then used to generate more samples and so to train a probabilistic model. 
As explained is Section~\ref{subsection:kde}, we use the generated data only for training 
probabilistic models but not for testing because of the considerable bias.  

Next, we employ the greedy EM algorithm to train bivariate GMMs based on the extracted distance and 
orientation features for each spatial relationship. The results are PDFs for each spatial relationship that, 
as the initial outset suggests, can be used to estimate the unknown position of spatial objects.

Our approach has been implemented in Matlab and all the experiments were conducted on an Intel(R) Core(TM) i5-2400 CPU at 3.10GHz with 8GB of RAM, running Ubuntu Linux 11.10.

\subsection{Visualization of Quantitative Spatial Relations} 
\label{sub:result_section_1}

The most important means of assessing the result is to visualize the quantified spatial relations. 
We divided the London bounding box to filter the input data by means of a $50\times50$ 
spatial grid. Each grid cell corresponds to a $4.4km \times 2.2km$ spatial 
extent (Longitude, Latitude). Given two spatial objects and the known location at the center of the grid, 
we plot for each grid cell the positional probability of the unknown location, i.e., how likely it would be for the unknown 
spatial object to be located in a specific grid cell.
Using a heat map, warmer colors (red) indicate higher probabilities.

\begin{figure*}[t]
\begin{center}
%\begin{tabular}	{c c c c}
\subfigure[North]{\includegraphics[width=2.2in]{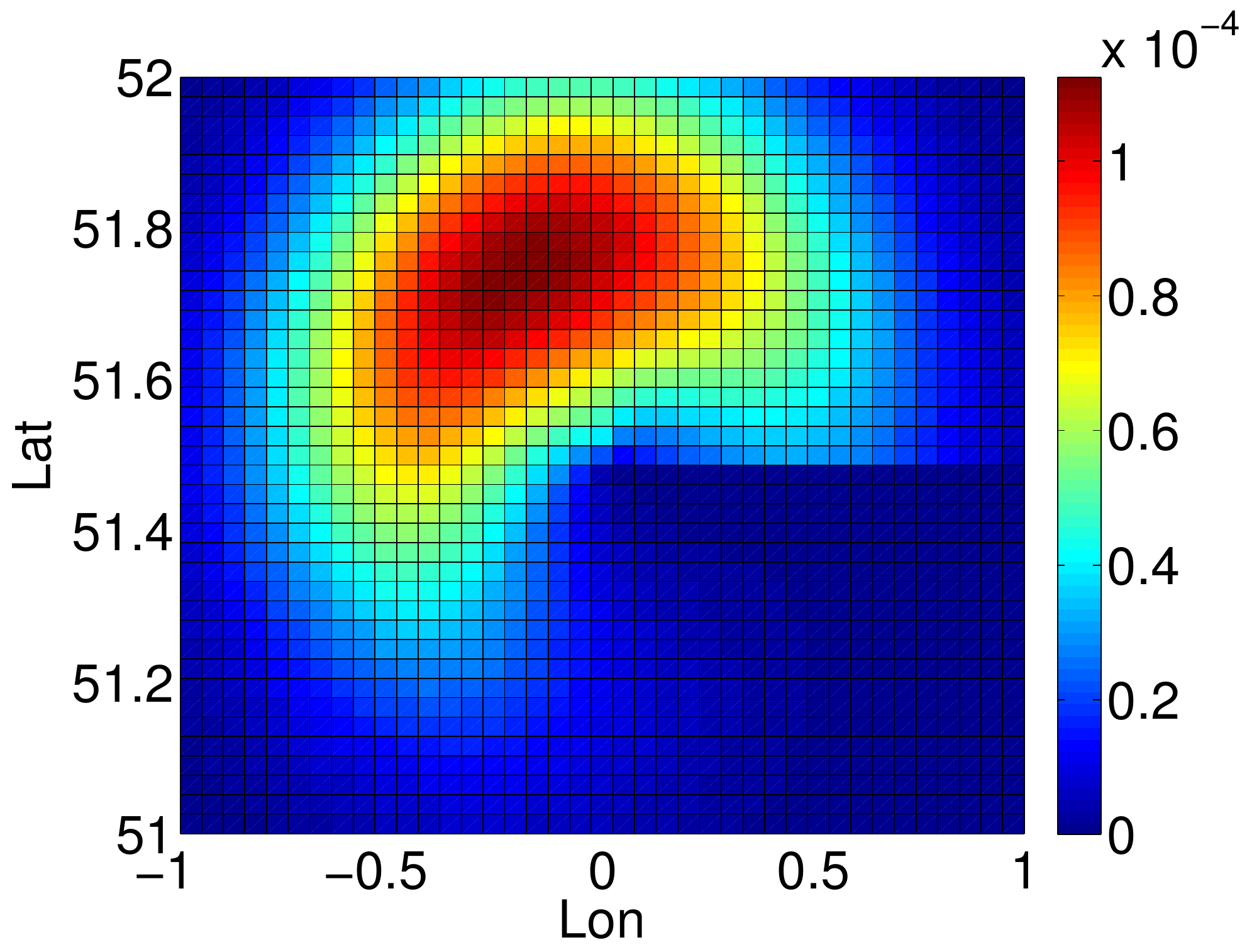}\label{subfig:uncorelviz_a}}
\subfigure[South]{\includegraphics[width=2.2in]{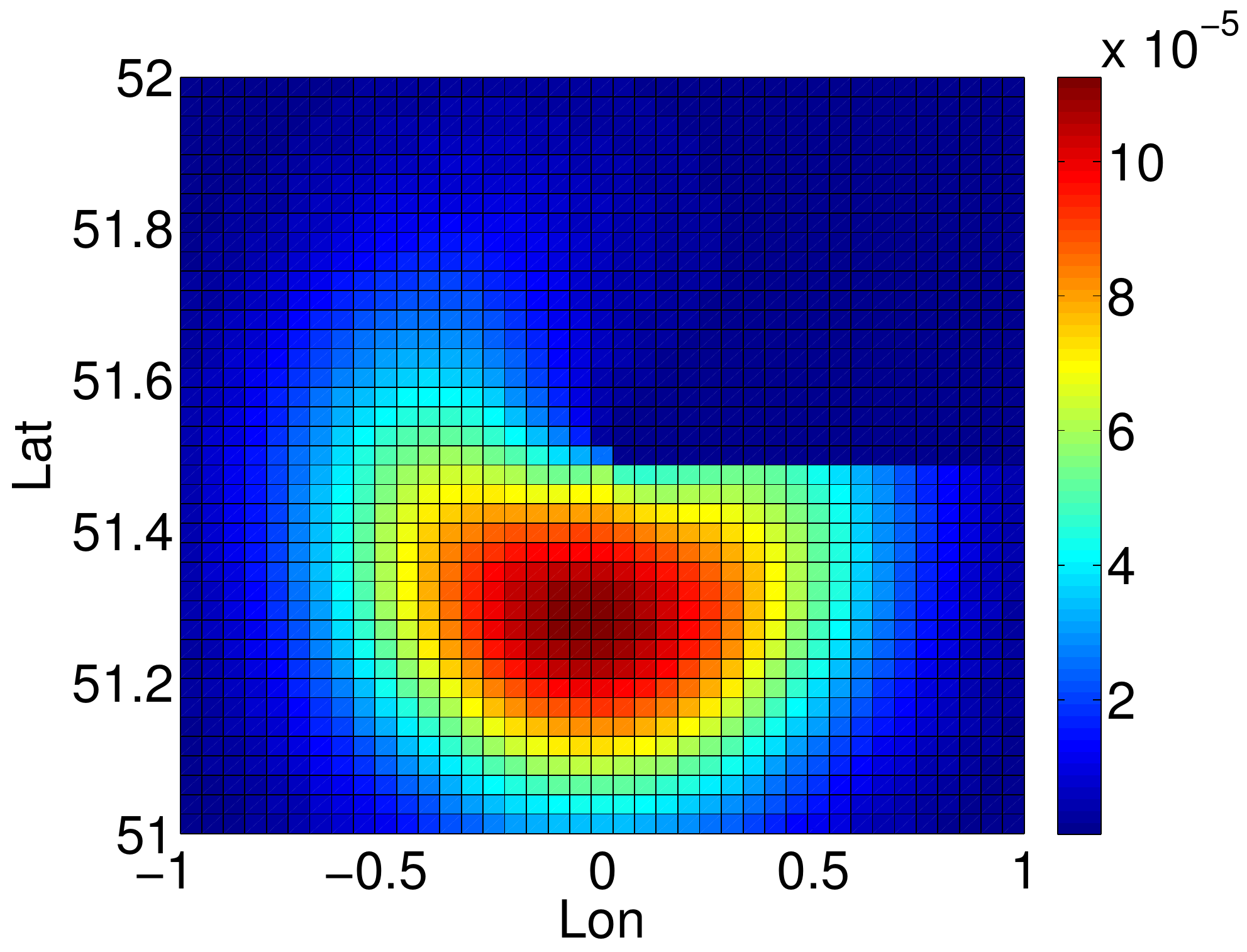}\label{subfig:uncorelviz_b}}\\
\subfigure[Near]{\includegraphics[width=2.2in]{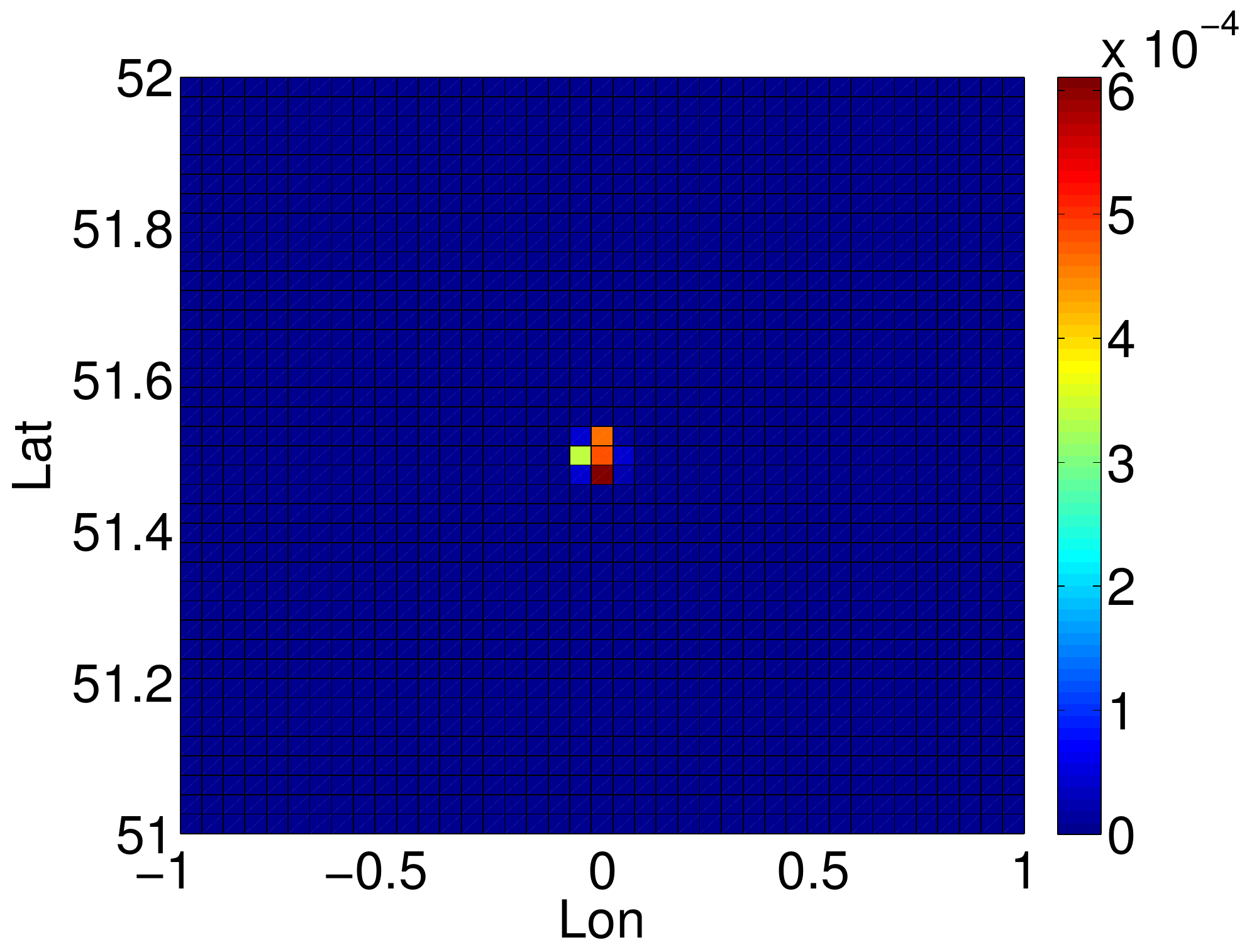}\label{subfig:uncorelviz_c}}  
\subfigure[In]{\includegraphics[width=2.2in]{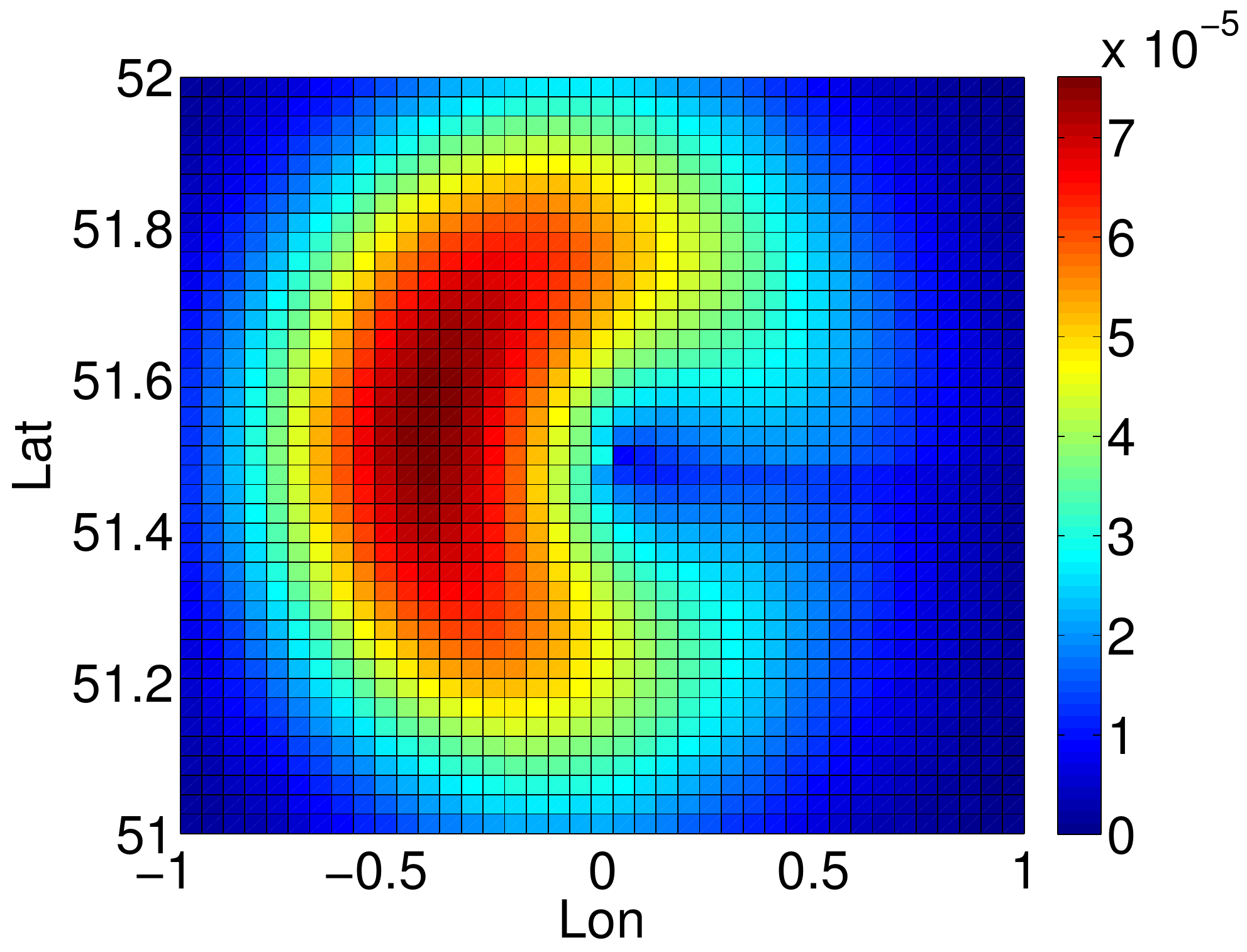} \label{subfig:uncorelviz_d}}
%(a) North & (b) South & (c) Near & (d) In \\
%\end{tabular}
\caption{Probabilistic heat maps for four basic spatial relationships: 1-Component Gaussian Mixture Models for the uncorrelated distance and orientation case. All figures illustrate 
the case where a POI is conntected with the center of the grid, with the respective spatial relation.} 
\label{fig:uncorelviz}
\end{center}
\end{figure*}

Figure~\ref{fig:uncorelviz} shows four spatial relationships modeled as one-component GMMs, with distance and orientation considered as uncorrelated random variables. 

\begin{figure*}[t]
\begin{center}
%\begin{tabular}{c c c c} 
\subfigure[Max 1-Component Uncorrelated Case]{\includegraphics[width=2.2in]{drawings/NorthMax1CompUncorel.pdf}\label{subfig:uncorelviz_north_a}}  
\subfigure[Max 1-Component Correlated Case]{\includegraphics[width=2.2in]{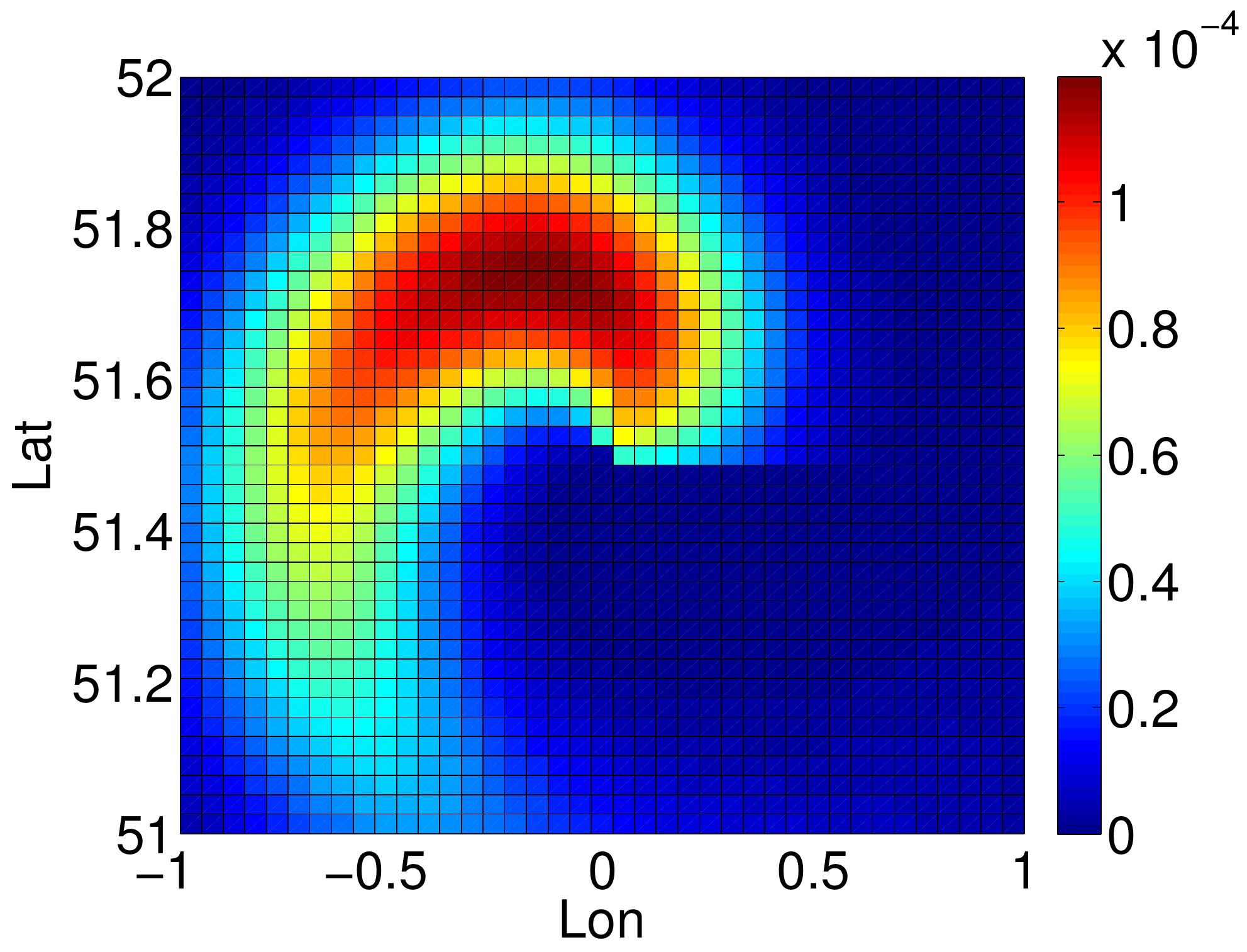}\label{subfig:corelviz_north_b}}\\
\subfigure[Max 5-Component Uncorrelated Case]{\includegraphics[width=2.2in]{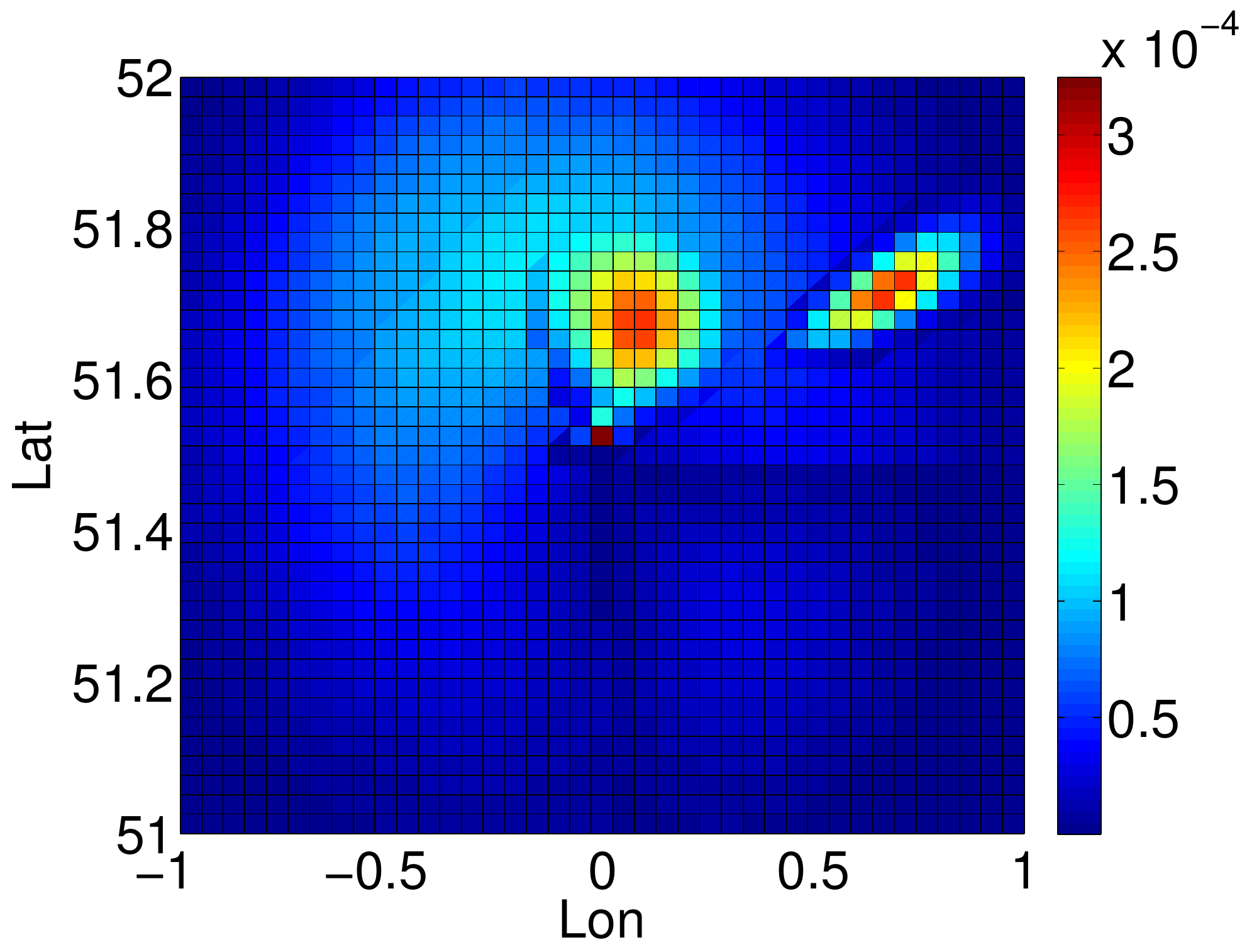}\label{subfig:uncorelviz_north_c}} 
\subfigure[Max 5-Component Correlated Case]{\includegraphics[width=2.2in]{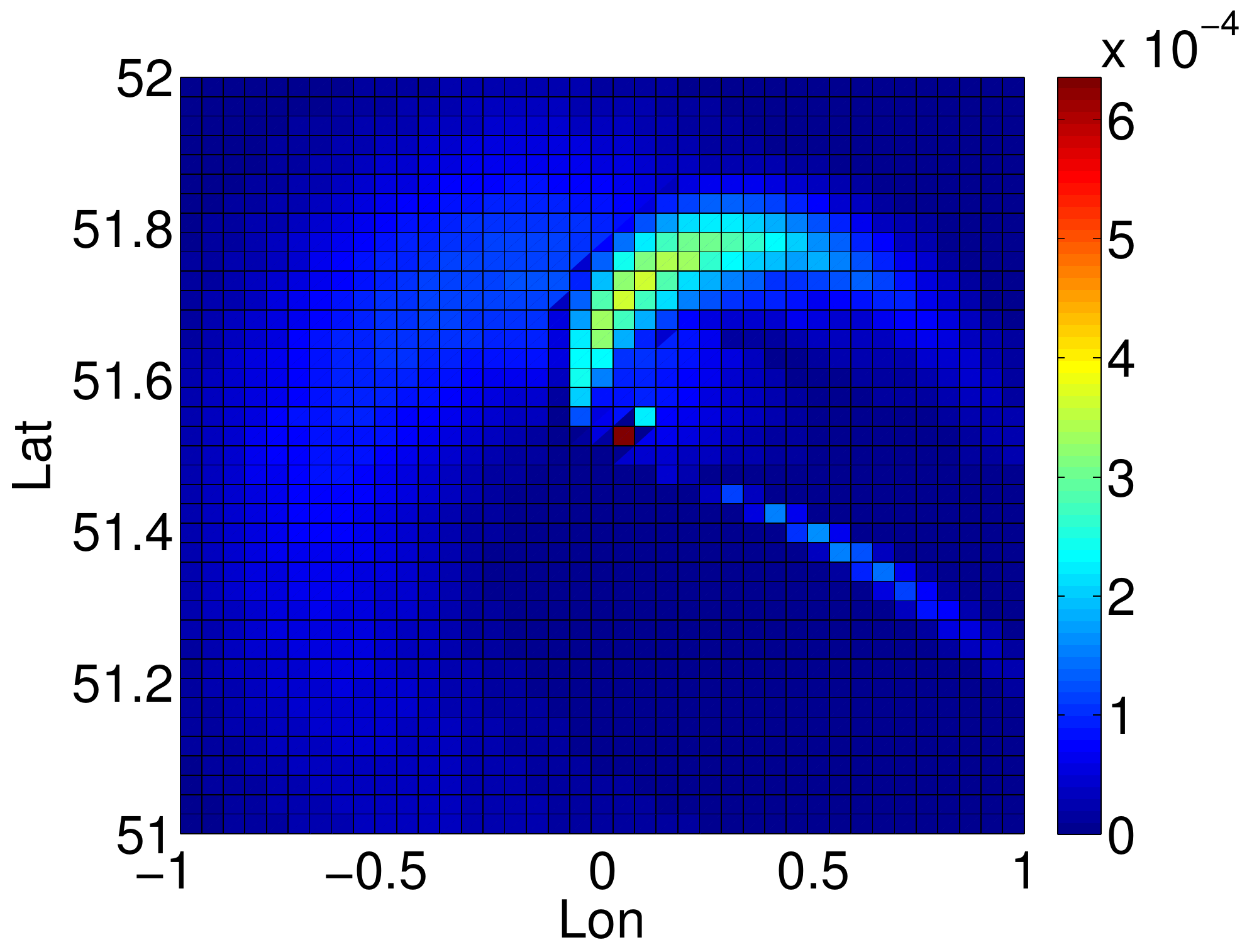}\label{subfig:corelviz_north_d}}  
%North Max1-Comp & North Max5-Comp & North Max10-Comp & North Max20-Comp\\
%(a) Max 1-Component & (b) Max 1-Component & (c) Max 5-Components & (d) Max 5-Components\\
%\end{tabular}
\caption{Probabilistic heat maps for \emph{North}: (a), (b) show correlated and uncorrelated distance and orientation case for a max of 1 Gaussian component per GMM while  (c), (d) 
show correlated and uncorrelated distance and orientation case for a max of 5 Gaussian components per GMM.}
\label{fig:north} 
\end{center}
\end{figure*}

The proposed modeling based on distance and orientation features performs especially well in some of the cases. More specifically, for the cases of \emph{North} (cf. Figure~\ref{subfig:uncorelviz_a}), 
\emph{South} (cf. Figure~\ref{subfig:uncorelviz_b}) and \emph{Near} (cf. Figure~\ref{subfig:uncorelviz_c}) the proposed model returns high probabilities in the expected regions.
On the other hand, the case of \emph{In} (cf. Figure~\ref{subfig:uncorelviz_d}) seems to include a lot of statistical noise due to the general uncertain nature of user generated content. For example, high distance and orientation variance values for the cases of 
\emph{On} and \emph{In} are caused by the the fact that most of the sentences that contain these spatial relations are of 
the form \emph{\textbf{POI} in London} and \emph{\textbf{POI} on river Thames}.     

\subsubsection{\textbf{Optimal Number of Gaussian Components}}

An important parameter when generating GMMs is the maximum number of Gaussian components. 
Such a limit is simply a stop criterion in the GMM training process and does not mean that the final component will converge to this upper limit, e.g., 
it might already converge to a lower number of components. 
Figure~\ref{fig:north} illustrates the case of spatial relation \emph{North}. The heat maps of Figures~\ref{subfig:uncorelviz_north_a} and \ref{subfig:corelviz_north_b} show the cases of a maximum of 1 Gaussian components per mixture when
distance and orientation are considered as uncorrelated and correlated, respectively.
The heat maps of Figures~\ref{subfig:uncorelviz_north_c} and \ref{subfig:corelviz_north_d} show the cases of a maximum of 5 Gaussian components per mixture, when distance and orientation are considered as uncorrelated and correlated, respectively.

For both uncorrelated and correlated cases, Figures~\ref{subfig:uncorelviz_north_c} and \ref{subfig:corelviz_north_d} show that by stepwise increasing the maximum number 
of Gaussian components, high probabilities tend to accumulate in fragmented small regions. 
The reason is that higher number of mixtures per GMM leads to components that are converging on their parameters  (mean, covariance, component weight) 
based on more dense regions of the dataset, e.g., regions with more data samples will become dominant components. The weight of such \emph{dominant components} 
(it is denoted as $w$ in Section~\ref{subsection:gmmem}) is higher than the weight of other components in the final GMM. 
This results in fragmented high probability regions in the final heat maps. 

As a result of this phenomenon, a major question is to the best approach to decide about the number of components per GMM. 
From an intuitive point of view, a smaller number of Gaussian components performs better as it preserves spatial generality, i.e., trends. However, 
as high probability regions are larger, they might result in inefficient location prediction tasks. From a statistical point of view, a high number of mixture 
components results in more accurate probabilistic models and better classification performance in most of the cases.
Unfortunately, the latter approach leads to sparse and small, high probability regions, which could be characterized as biased to the specific characteristics of 
the geographic region from where the dataset is taken (London in our case). 

\begin{figure*}[here!]
\begin{center}
\subfigure[]{\includegraphics[width=5.5cm,height=4.8cm]{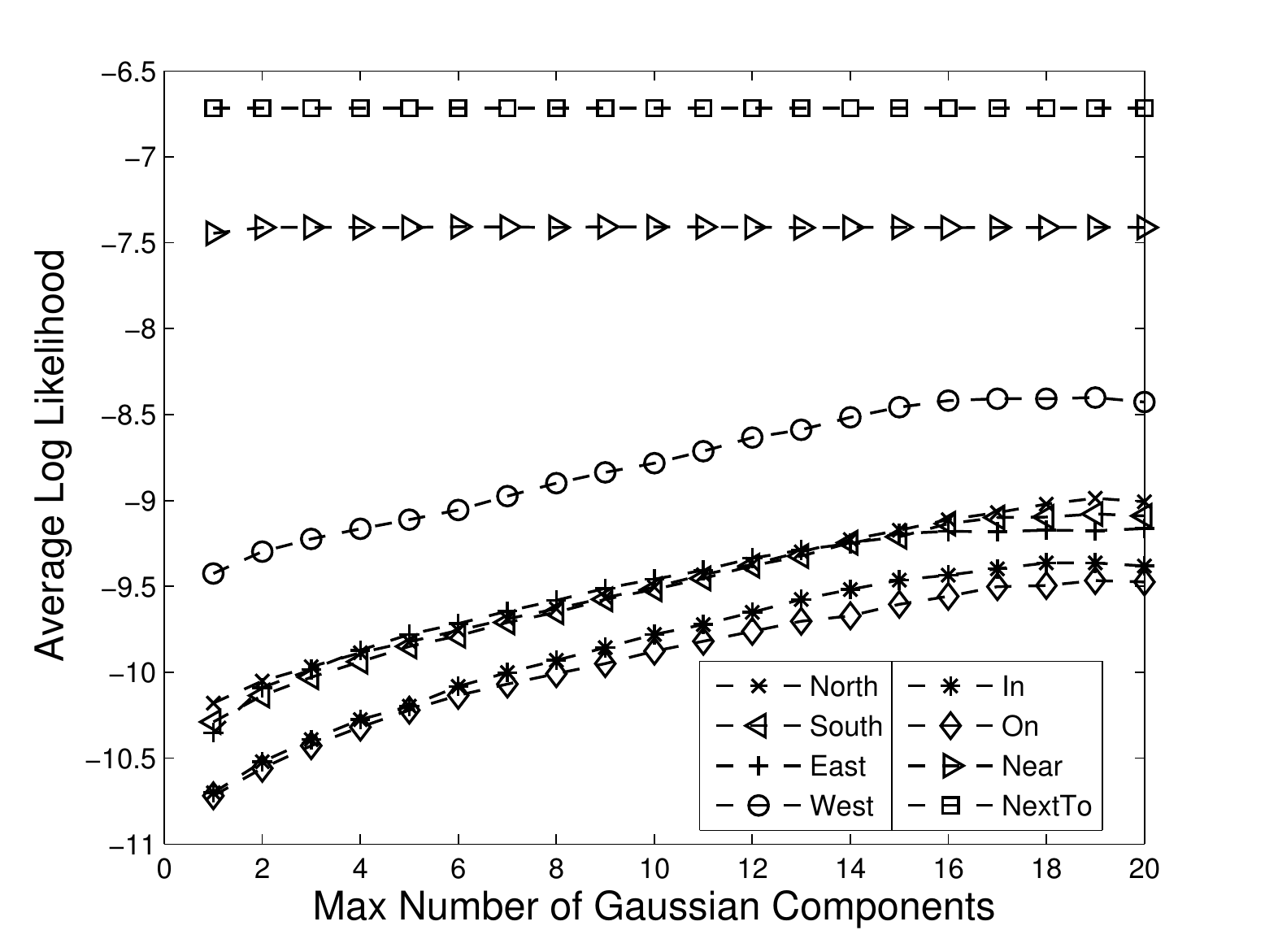} \label{subfig:llkl_a}} 
\subfigure[]{\includegraphics[width=5.5cm,height=4.8cm]{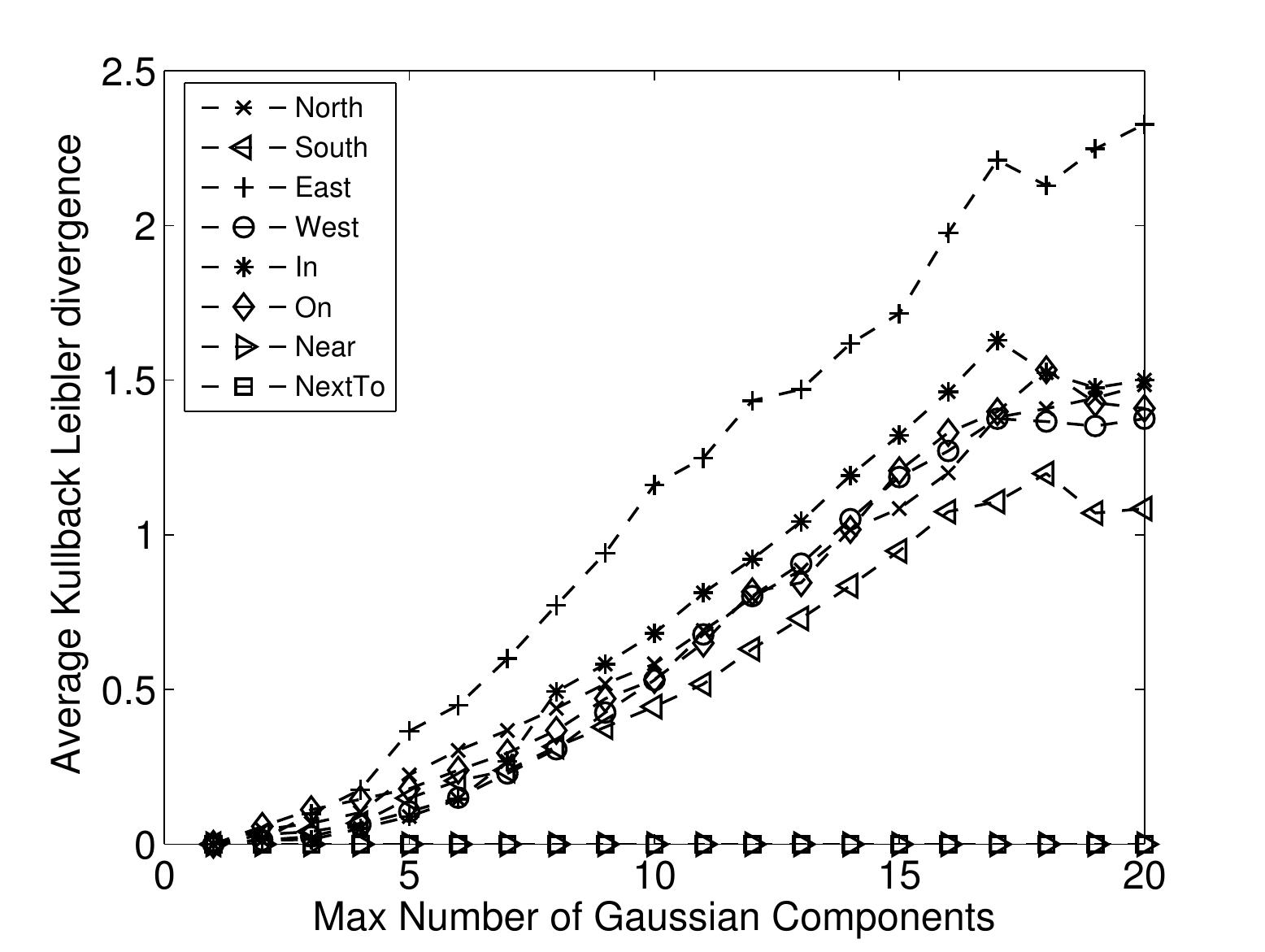} \label{subfig:llkl_b}} \\
\subfigure[]{\includegraphics[width=5.5cm,height=4.8cm]{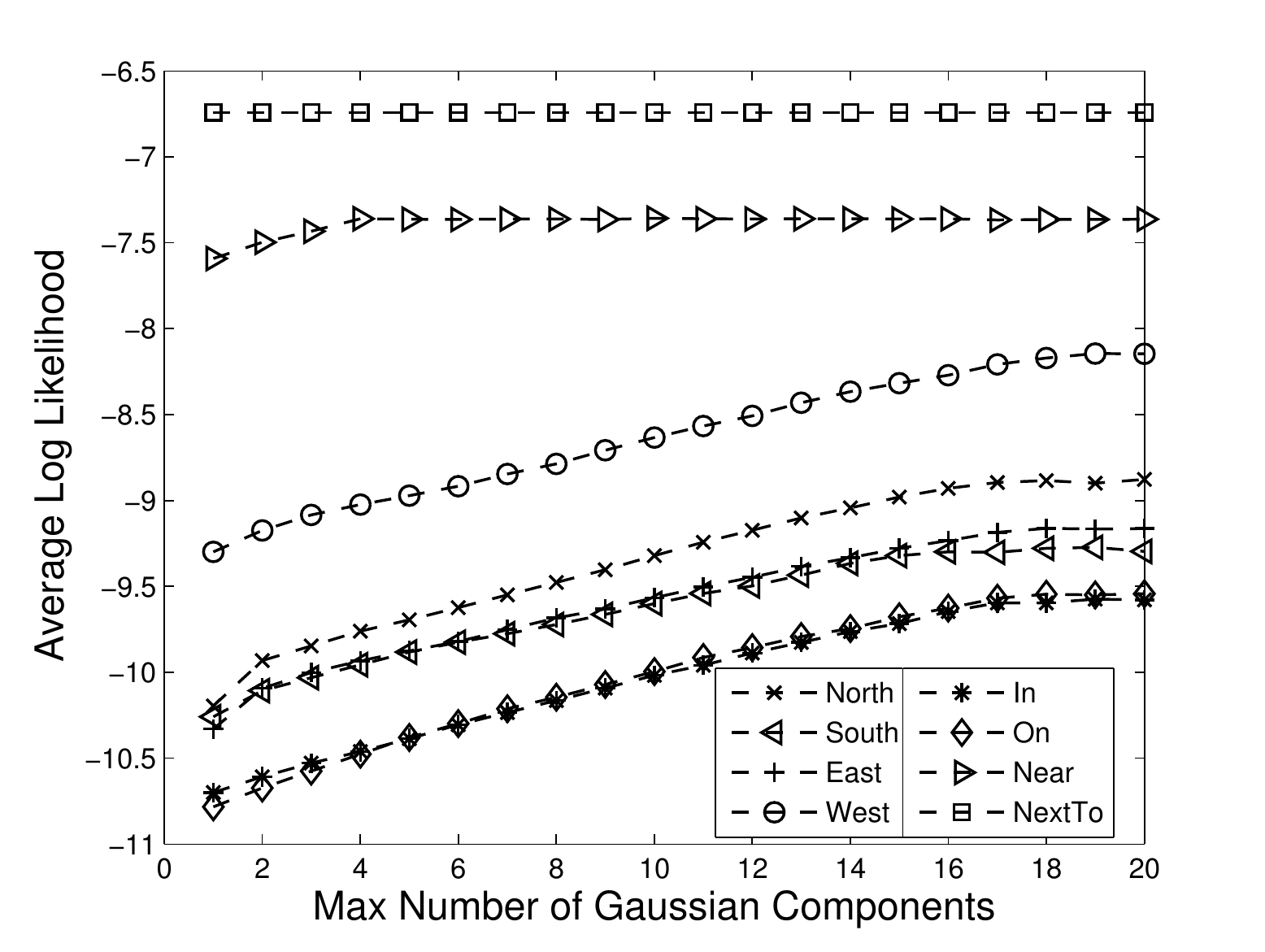} \label{subfig:llkl_d}} 
\subfigure[]{\includegraphics[width=5.5cm,height=4.8cm]{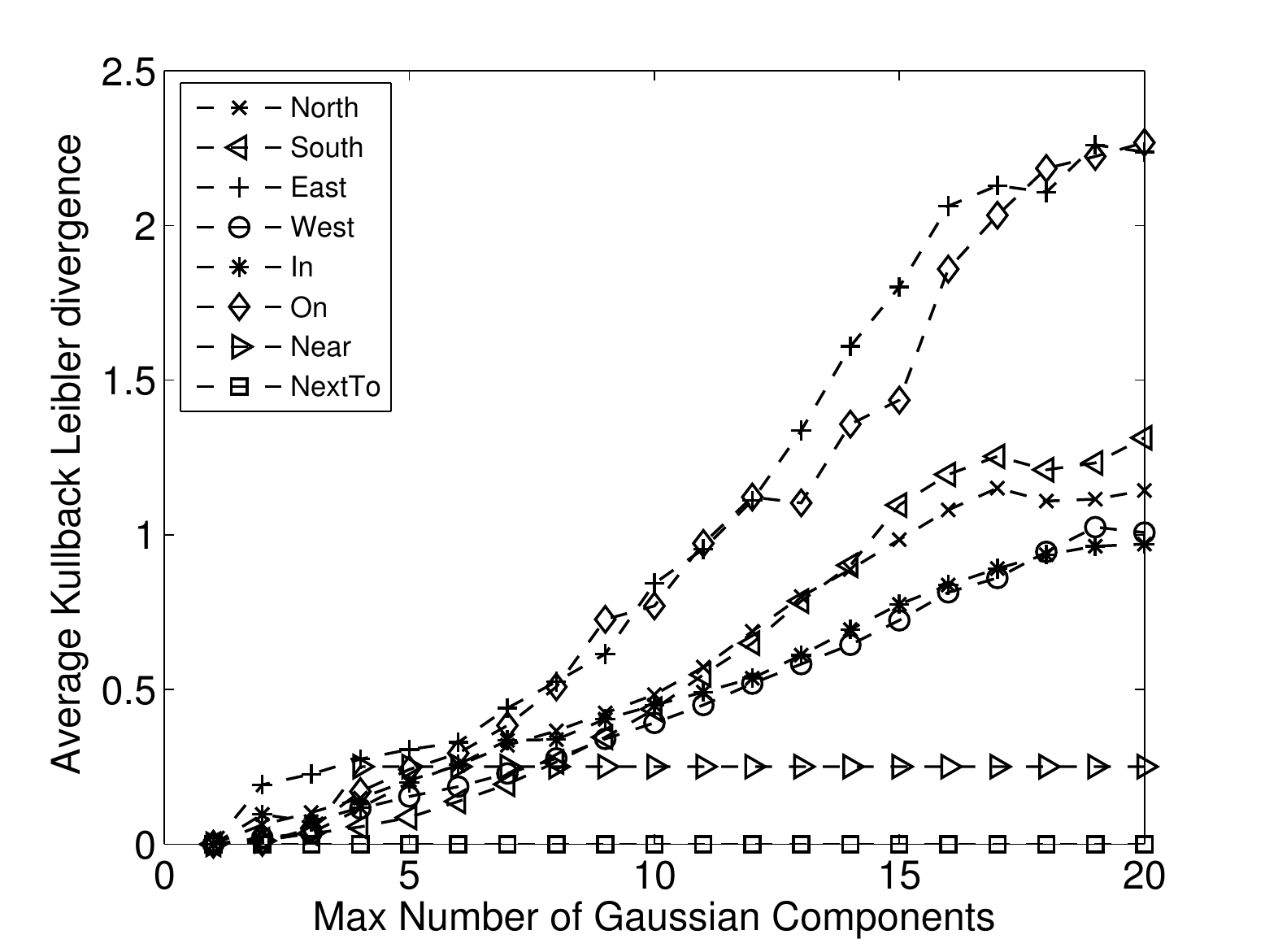} \label{subfig:llkl_e}} \\
\subfigure[]{\includegraphics[width=5.5cm,height=4.8cm]{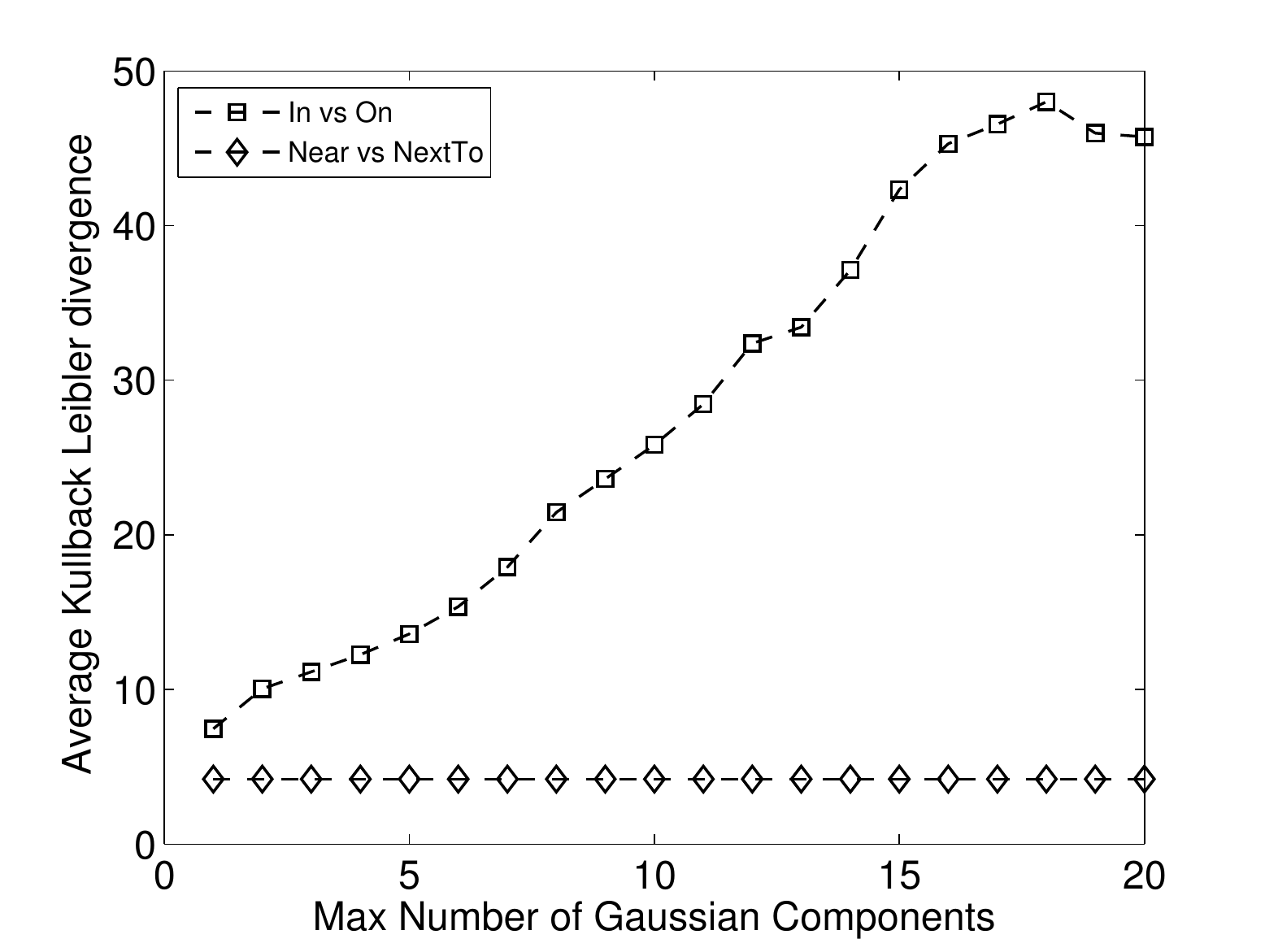} \label{subfig:llkl_c}} 
\subfigure[]{\includegraphics[width=5.5cm,height=4.8cm]{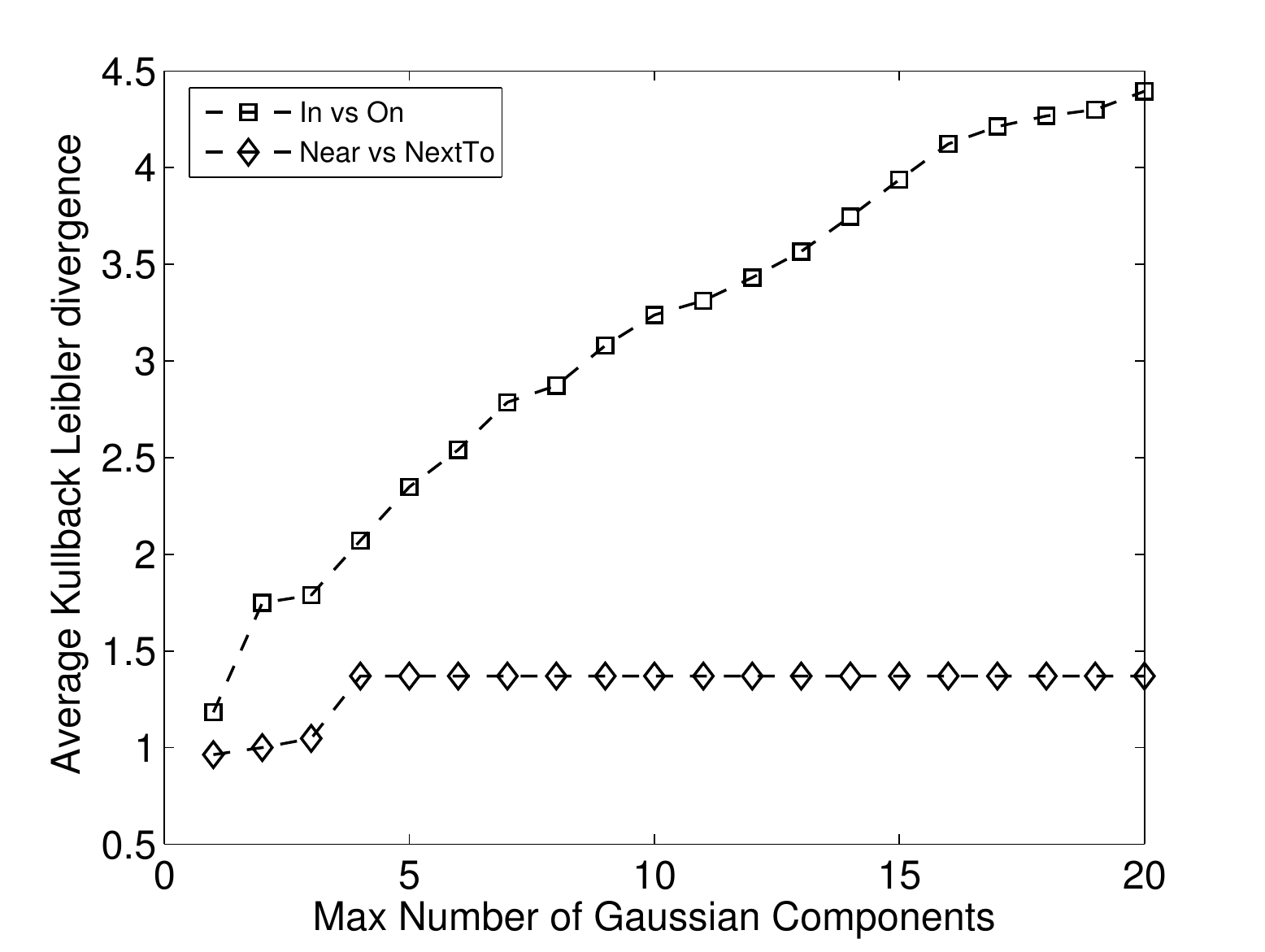} \label{subfig:llkl_f}} 
\caption{(a), (c) Average log-likelihood vs maximum number of Gaussian components for correlated and uncorrelated distance and orientation case respectively.  
(b), (d) Average KL divergence between the baseline 1-component GMM and the final converged GMM after each step of 
increasing the maximum number of components for correlated and uncorrelated distance and orientation case respectively. (e), (f) Average KL diverge between spatial relationship pairs ``In-On'' and ``Near-Nextto'' 
for correlated and uncorrelated distance and orientation case respectively.}
\label{fig:llkl} 
\end{center}
\end{figure*}

In Figures~\ref{subfig:llkl_a} and ~\ref{subfig:llkl_d}, we depict the \emph{average log-likelihood}, e.g., we estimated parameters and log-likelihoods for each spatial relation model 
running the greedy learning algorithm 100 times per maximum component step and stepwise increasing the maximum Gaussian components per GMM. Figures~\ref{subfig:llkl_a}
and \ref{subfig:llkl_d} show the cases of correlated and uncorrelated distance and orientation random variables, respectively. 
In both cases, most of the spatial relation models converge on a high number of components, i.e., 16-17. 
Only spatial relations \emph{Near} and \emph{NextTo} converge on a smaller number of components. This means that statistically, most of the spatial relationships 
should be modeled with an upper limit of Gaussian components close to 16 or 17. In practice, this will result in fragmented spatial probabilities (heat maps) as outlined above.

Concluding, we realize that based on the log-likelihood measurements, there are statistically correct and sometimes optimal solutions for deciding the number of components. 
The difficult part in our case is the balance between statistical and intuitive robustness. 

Based on a user generated dataset, we believe that 
GMMs with a number of components between 1 and 10 are statistically correct (but not optimal) and intuitively efficient to model spatial relations.

\subsubsection{\textbf{Correlated vs. Uncorrelated Features}}

Correlation between distance and orientation is another important issue when training GMMs. 
Literature suggests that most of the classification approaches perform better when probabilistic models are trained taking into consideration the correlation between random variables. 
In our work, visualization shows that there is a high correlation between distance and orientation for some but not all cases. Figure~\ref{fig:north} illustrates the case of \emph{North}. 
For the heat maps shown in Figures~\ref{subfig:uncorelviz_north_a} and \ref{subfig:uncorelviz_north_c}, distance and orientation are considered uncorrelated, for  Figures~\ref{subfig:corelviz_north_b} and \ref{subfig:corelviz_north_d}, 
they are considered as correlated random variables. 
Intuition suggests that we can not guarantee that the correlated case performs better, even if we are sure that distance and orientation are correlated. 
The \emph{North} case should result in high probabilities for the top part of the grid as it should be the case for all directional relations. 
However, based on visual results and heat maps for all modeled spatial relations, we observe that distance and orientation seem to be less 
correlated for the cases of \emph{In} (cf. Figure~\ref{subfig:uncorelviz_d}) and \emph{On}, and tend to have zero correlation, e.g., region around the center of 
the grid with almost equal probabilities, for the cases of \emph{Near} (cf. Figure~\ref{subfig:uncorelviz_c}) and \emph{NextTo}. 
As expected, this leads as to the conclusion that some spatial relations are independent of orientation, e.g., only distance could model them efficiently. 
This also means that distance and orientation should be modeled as independent random variables. 

Summing up, based on user-generated content, we believe that directional relations like \emph{North} should be modeled taking correlation between distance and orientation into consideration. 
On the other hand, topological relations such as \emph{In} and metric relations like \emph{Near} tend to be independent of orientation, which means that correlation between distance and 
orientation should not be taken into consideration during modeling. 

\subsection{Similarity Between Quantified Spatial Relations} 
\label{sub:result_section_2}

Besides a visual inspection, it is important to have a quality metric to assess the probabilistic spatial relation quantification. 
We use Kullback-Leibler (KL) divergence (i) to assess the similarity between converged GMMs of the same relation
%, as we stepwise increase the maximum number of components that each converged GMM could have
and (ii) to measure the similarity between some spatial relationships that tend to follow similar patterns. 

Figures~\ref{subfig:llkl_b} and \ref{subfig:llkl_e} illustrate the KL divergence between the baseline 1-component GMM and the final converged GMM after each step of 
increasing the maximum number of components for correlated and uncorrelated cases, respectively. 
Most of the models tend to diverge from the baseline model as we increase the maximum number of components. Only the models for \emph{Near} and \emph{NextTo} 
have low and zero distance from their baseline model. This matches the corresponding log-likelihoods illustrated in Figures~\ref{subfig:llkl_a} and \ref{subfig:llkl_d}, which remain almost stable. 
In these examples, with a small number of Gaussian components for \emph{Near} and one Gaussian component for \emph{NextTo} the log-likelihoods remains stable.  

Finally, Figures~\ref{subfig:llkl_c} and \ref{subfig:llkl_f} show that spatial relation pairs \emph{Near-NextTo} and \emph{On-In} exhibit similar characteristics. 
To assess this similarity, we measured the KL divergence for all cases of their models. The aforementioned figures show that the pair \emph{On-In} seems to diverge 
as we increase the number of components for both correlated and uncorrelated cases. However, the pair \emph{Near-NextTo} exhibits low values of KL divergence for all cases.
This leads to the conclusion that people use more than one language expression to describe the same spatial relation. For our examples, this means that we could merge the 
cases of \emph{Near} and \emph{NextTo} into one probabilistic model.\\

% \subsection{Summary} 
% \label{sub:summary}
% 
% With the objective to quantify qualitative spatial knowledge in the form of spatial relationships, this work proposed Gaussian Mixture Models as a means to 
% derive probability distributions for specific relationship cases. The experiments with data derived from travel blogs and limited to the geographic area of London showed that 
% given sufficient data it is indeed possible to quantify such relations, i.e., how far is \emph{Far}. Visualizations showed the actual spatial extent of spatial relations. In addition, we use KL divergence
% to compare the derived GMMs during various stage of the process and to compare GMMs of different relations to see whether they actual express the same geospatial fact.

%!TEX root = paper.tex

\section{Conclusions} 
\label{sec:conclusions}

The increase in available user-generated data provides a unique opportunity for the generation of rich datasets in geographical information science.
In this work, we provide a quantitative approach for the representation of qualitative spatial relations extracted from such data based on training probabilistic models. 
The proposed scheme returns estimates of uncertain object locations based on distance and orientation features as provided by human reporters in relation to  
known object locations. To achieve these desiderata%plural word
, we propose a greedy learning algorithm based on the Expectation Maximization (EM) framework to train probabilistic models 
over spatial relationships; here, we restrict our attention on GMM models. %The results are visualized using a spatial grid for indicating probable object locations. 
The proposed approach seems to be promising in terms of 
accurately capturing and representing spatial relationships.
%efficiency, with attractive results for the representation of spatial relationships. 
Distance and orientation features tend to describe all spatial relations that were extracted from user generated texts in an informative way. Moreover, our probabilistic approach seems to be robust in handling 
any uncertainties, which characterize observations in crowd-sourced text data. 
As a future research direction, we already have been investigating new NLP techniques for the automatic 
extraction of POIs and spatial relationship information from texts and we are very close to a practical and theoretically robust solution. This will enable us to evaluate additional
probabilistic and deterministic modeling techniques 
%for the quantitative representation of spatial relationships 
and to develop efficient text-to-map applications.

\section*{Acknowledgements} 
\label{sec:acknowledgements} 
%Acknowledge funding, i.e., research projects and input from others that are not co-authors of this paper but have somehow helped. 
The research leading to these results has received funding from the 
European Union Seventh Framework Programme - Marie Curie Actions, Initial Training Network GEOCROWD (http://www.geocrowd.eu) 
under grant agreement No. FP7-PEOPLE-2010-ITN-264994.
% section acknowledgements (end)
\bibliographystyle{abbrv}
\bibliography{geoinfo}

\end{document}